\documentclass[aps,prr,reprint,twocolumn,superscriptaddress,floatfix,nofootinbib,amsmath, amssymb, longbibliography]{revtex4-2}

\usepackage[utf8]{inputenc}
\usepackage{graphicx}
\usepackage{dcolumn} 
\usepackage{bm}
\usepackage{color}
\usepackage[normalem]{ulem}

\usepackage[english]{babel}
\usepackage{amsmath}
\usepackage{color}
\usepackage{graphicx}
\usepackage{physics}
\usepackage{ulem}
\usepackage{amsfonts}
\usepackage[export]{adjustbox}[2011/08/13]
\usepackage{setspace}

\usepackage{amssymb}
\usepackage{float}
\usepackage[caption=false]{subfig}
\usepackage{xr-hyper}
\usepackage{hyperref}

\makeatletter
\newcommand*{\addFileDependency}[1]{
\typeout{(#1)}%
\@addtofilelist{#1}
\IfFileExists{#1}{}{\typeout{No file #1.}}
}\makeatother

\newcommand*{\myexternaldocument}[1]{%
\externaldocument{#1}%
\addFileDependency{#1.tex}%
\addFileDependency{#1.aux}%
}
\myexternaldocument{./SM}
\newcommand{\q}[3][q]{#1^\pqty{#2}_{#3}}

\newcommand{\PF}[1]{\color{orange} \color{black}}

\newcommand{\be}{\begin{eqnarray}}
\newcommand{\ee}{\end{eqnarray}}

\begin{document}
\title{Protein Design by Integrating Machine Learning with Quantum Annealing and Quantum-inspired Optimization}
\author{Veronica Panizza}
\affiliation{Pitaevskii BEC Center,  Physics Department, Trento University, Via Sommarive 14, 38123 Povo (Trento), Italy.}
\affiliation{INFN-TIFPA, Via Sommarive 14, 38123 Povo (Trento), Italy.}
\author{Philipp Hauke}
\email{philipp.hauke@unitn.it}
\affiliation{Pitaevskii BEC Center,  Physics Department, Trento University, Via Sommarive 14, 38123 Povo (Trento), Italy.}
\affiliation{INFN-TIFPA, Via Sommarive 14, 38123 Povo (Trento), Italy.}
\author{Cristian Micheletti}
\affiliation{Scuola Internazionale Superiore di Studi Avanzati (SISSA),Via Bonomea 265, I-34136 Trieste, Italy.}
\email{cristian.micheletti@sissa.it}
\author{Pietro Faccioli}
\affiliation{Department of Physics University of Milano-Bicocca and INFN, Piazza della Scienza 3, I-20126 Milan, Italy.}
\email{pietro.faccioli@unimib.it}

\begin{abstract}
The protein design problem involves finding polypeptide sequences folding into a given three-dimensional structure. Its rigorous algorithmic solution is computationally demanding, involving a nested search in sequence and structure spaces. Structure searches can now be bypassed thanks to recent machine learning breakthroughs, which have enabled accurate and rapid structure predictions. Similarly, sequence searches might be entirely transformed by the advent of quantum annealing machines and by the required new encodings of the search problem, which could be performative even on classical machines.
In this work, we introduce a general protein design scheme where algorithmic and technological advancements in machine learning and quantum-inspired algorithms can be integrated, and an optimal physics-based scoring function is iteratively learned. In this first proof-of-concept application, we apply the iterative method to a lattice protein model amenable to exhaustive benchmarks, finding that it can rapidly learn a physics-based scoring function and achieve promising design performances. Strikingly, our quantum-inspired reformulation outperforms conventional sequence optimization even when adopted on classical machines. The scheme is general and can easily extended, e.g., to encompass off-lattice models, and it can integrate progress on various computational platforms, thus representing a new paradigm approach for protein design.
\end{abstract}
\maketitle

\section{Introduction}
In contrast to random polypeptide chains, most naturally occurring proteins fold rapidly and reversibly into a unique conformation that is solely determined by the sequence of amino acids, called the native state.~\cite{Lesk2004,Dill_Chan_1997,doi:10.1126/science.1208351}. This property is consistent with the interpretation that the native state typically corresponds to the free-energy minimum of the peptide chain~\cite{Anfinsen_1973,Anfinsen1975} and is kinetically accessible from generic conformers of the unfolded ensemble~\cite{levinthal1968there,Levinthal1969,Dill_Chan_1997,doi:10.1126/science.1208351}.
The unique thermodynamic properties of proteins and protein-like systems promoted by natural or artificial selection~\cite{Anfinsen_1973,Anfinsen1975,bryngelson1989intermediates,sali_shakhnovich_karplus_1994,li_emergence_1996,PhysRevLett.77.5433,Dill_Chan_1997,klimov_cooperativity_1998,chan_cooperativity_2004,fitzkee_reassessing_2004,ONUCHIC200470,doi:10.1126/science.1208351,dill2008protein,cocco2018inverse,negri_native_2021}, have long posed two fundamental challenges: (i) predicting protein structures given the chemical sequence and (ii) finding sequences that can fold onto a given target structure. These are known as protein folding and protein design problems, respectively. Because of their close connection, they are also called the \emph{direct} and the \emph{inverse} protein folding problems.

From a thermodynamic perspective, both challenges can be fully specified by defining the (effective) energy $E(\Gamma, S)$  of a polypeptide chain as a function of its sequence, $S$, and its conformational state,  $\Gamma$. By effective energy, we intend that $E$ includes contributions from the thermodynamic integration of the solvent degrees of freedom. Solving the direct protein folding problem for a given polypeptide sequence $S$ involves finding the conformer(s) $\Gamma$ with the largest occupation probability in canonical equilibrium,
\begin{equation}
P(\Gamma | S) = \frac{e^{-\beta E\pqty{\Gamma,S}}}{\sum_{\Gamma'} e^{-\beta E\pqty{\Gamma',S}}}\ ,
\label{eqn:des}
\end{equation}

\noindent where $\beta$ is the inverse thermal energy in physiological conditions and the sum in the denominator runs over the possible conformational states. Foldable polypeptide chains, such as naturally-occurring proteins, are characterized by the thermodynamic stability of the state $\Gamma$ maximizing Eq.~\eqref{eqn:des}, i.e., $P(\Gamma | S) > p_\mathrm{fold}$, where $p_\mathrm{fold}$ is a suitable threshold, typically larger than 0.5. In this case, $\Gamma$ is the native state of $S$.

Conversely, solving the inverse folding problem for a given target state $\Gamma_T$ amounts to finding a sequence $S$, if any exists, such that
\be
P(S|\Gamma_T ) &=&
 \frac{e^{-\beta E\pqty{\Gamma_T,S}}}{\sum_{\Gamma'} e^{-\beta E\pqty{\Gamma',S}}}\nonumber\\
 &\equiv& e^{-\beta (E\pqty{\Gamma_T, S} - F(S))}\ge p_\mathrm{fold}\, ,
\label{eqn:des2}
\ee
where $F(S)= -\frac{1}{\beta}\ln \sum_{\Gamma'} e^{-\beta E(\Gamma',S)}$ is the free energy of sequence $S$.
Sequences that satisfy inequality (\ref{eqn:des2}) are said to design the target state $\Gamma_T$~\cite{PhysRevLett.76.323,seno_optimal_1996}.

Thus, solving the design problem is equivalent to finding sequences that minimize the function
\begin{equation}
G(S) = E(\Gamma_T, S) - F(S)  \ ,
\label{eq:exact_scoring_function}
\end{equation}
and satisfy the inequality in Eq.~(\ref{eqn:des2}).

A key point is that the computational demands for solving the above rigorous physics-based formulations of the direct and inverse folding problems differ greatly.
Solving the former involves, in principle, computing the energy of the sequence of interest over the entire set of physically viable conformational states. Conversely, to solve the design problem, it is not sufficient to compute the energy of all viable sequences on the given target structure, $\Gamma_T$. Indeed, the sequences that minimize the energy when mounted on $\Gamma_T$ may fold into a different structure, $\Gamma^\prime$, with even lower effective energy, i.e., $E(\Gamma^\prime, S)  < E(\Gamma_T, S)$~\cite{PhysRevLett.76.323,seno_optimal_1996}.
Hence, solving the design problem involves two nested searches: one over the sequences and one over the structures~\cite{PhysRevLett.76.323,seno_optimal_1996,micheletti1998protein,irback_monte_1998}. For this reason, much effort has been spent on finding practical schemes and approximations to curb the computational expenditure entailed by this problem~\cite{shakhnovich_protein_1998,PhysRevLett.76.323,seno_optimal_1996,micheletti1998protein,irback_monte_1998,huang_coming_2016,korendovych_novo_2020}.

A novel twist in this direction has come from recent advancements in machine learning~\cite{Paladino_Marchetti_Rinaldi_Colombo_2017,Goverde_Wolf_Khakzad_Rosset_Correia_2022,Jendrusch_Korbel_Sadiq_2021}. Relevant examples include the development of Bayesian learning design strategies~\cite{Takahashi_Chikenji_Tokita_2021}, and the experimental validation of deep-learning models~\cite{anishchenko_novo_2021}, including generative ones~\cite{watson_novo_2023,ingraham_illuminating_2023}.
On the one hand, these approaches provide elegant and valuable demonstrations of the striking extent to which sequence-structure correlations present in protein databases might be harnessed by empirical scoring functions for protein design.  In perspective, such endeavors could emulate the breakthrough in the empirical solution to the direct folding problem, where deep neural networks now yield remarkably reliable predictions~\cite{Jumper_Evans_Pritzel_et_2021}.

On the other hand, an inherent limitation of all such  empirical methods is the lack of interpretability. In contrast, physics-based approaches, based on an explicit definition of the energy function $E(\Gamma, S)$, would enable abstracting principles applicable to more general contexts~\cite{Shi_Jie_Chen2023}.
For this reason, the quest for computationally-amenable physics-based approaches to the protein design and related problems remains an active research avenue as well as a natural testbed for new computing hardware paradigms, including quantum computing~\cite{perdomo_truncik_tubert_brohman_rose_aspuru_guzik_2008,perdomo_ortiz_dickson_drew_brook_rose_aspuru_guzik_2012,robert_barkoutsos_woerner_tavernelli_2021,khatami_gate-based_2023,irback_knuthson_mohanty_peterson_2022}, as we shall discuss later.

In physics-based schemes, the cornerstone notion is that the energy function $E(\Gamma, S)$ in Eq.~\eqref{eq:exact_scoring_function} is the only theoretical ingredient needed to solve both the direct and the inverse folding problems. However, in detailed atomistic approaches, even a single computation of $E(\Gamma, S)$ would require extensive calculations, e.g., to integrate out the solvent degrees of freedom. Customarily, this prohibitively expensive computation is alleviated by resorting to coarse-grained models and implicit-solvent energy functions. At the same time, coarse-graining also tames the complexity of the sampling problem by smoothing the energy landscape and drastically reducing the number of conformational degrees of freedom~\cite{micheletti2000recurrent,kolodny2002small,camproux2004hidden,pandini2010structural,mackenzie2016tertiary,krupa2017maximum}. Yet, reliably estimating the functional form and the parametrization of the effective energy $E(\Gamma,S)$ remains challenging.

The main goal of the present study is to demonstrate that it is possible to integrate advancements in both machine learning and quantum computing technologies to tackle the design problem without abandoning the physics-based standpoint of Eqs.~(\ref{eqn:des2},\ref{eq:exact_scoring_function}). In this context, the research in quantum computing may also drive the development of radically new physics-based formulations that are advantageous even when implemented on classical machines \cite{slongo2023quantum}.


In recent years, several quantum-inspired algorithms have been proposed to unveil protein sequence-structure relationship ~\cite{perdomo_truncik_tubert_brohman_rose_aspuru_guzik_2008, perdomo_ortiz_dickson_drew_brook_rose_aspuru_guzik_2012, babbush_love_aspuru_guzik_2014, robert_barkoutsos_woerner_tavernelli_2021, wong_chang_2022,irback_knuthson_mohanty_peterson_2022},  compute protein folding pathways~\cite{hauke_mattiotti_faccioli_2021,ghamari_hauke_covino_faccioli_2022}, and more generally address equilibrium properties polymeric systems~\cite{micheletti_hauke_faccioli_2021,slongo2023quantum}.
By contrast, the protein design problem has been tackled by comparatively fewer attempts using algorithms designed for quantum hardware. Such pioneering efforts have relied on lattice protein models~\cite{Lau_Kit_Dill_1989,sali_shakhnovich_karplus_1994,yue_fiebig_thomas_chan_shakhnovich_dill_1995,li_emergence_1996}
because their discrete nature enables a straightforward mapping onto the quantum simulation hardware. In Ref.~\cite{khatami_gate-based_2023}, the authors used a gate-based quantum algorithm to reshuffle a sequence to minimize its energy on a reference structure. In contrast, in Ref.~\cite{Irback_Knuthson_Mohanty_Peterson_2024} a quantum-annealing platform was used for an analogous objective. Both studies employed a simplified two-letter amino acid alphabet and postulated the effective amino acid interactions.

In our first illustrative application, we, too, resort to minimalistic lattice protein models. This choice is particularly suited to assessing the accuracy of our scheme, allowing us to better control the sources of errors. Indeed, it eliminates the uncertainties associated with heuristic machine-learning algorithms for protein folding, as they can be replaced by an exhaustive search of the conformational space. Furthermore, it enables to assess the accuracy through which our iterative learning scheme is able to learn the underlying physics-based energy function.

Even after this major simplification, the combinatorial search over the sequence space can be computationally demanding.
Quantum annealing machines are ideally suited to solve this kind of discrete combinatorial optimizations after a suitable mathematical reformulation, or encoding, of the original problem. A very relevant question to address is if such reformulation can lead to performance improvement even when adopted on classical machines \cite{slongo2023quantum}.

We answer this question in the affirmative. Indeed, in our proof-of-concept study, quantum-encoded approaches implemented on both classical and quantum computers outperform a well-established scheme based on simulated annealing. At the same time, our iterative machine learning scheme enables us to reach solutions to the design problem with a high success rate.
Our algorithm's main merit is its ability to simultaneously harness the new possibilities offered by Machine Learning applications and promised by the advancements in quantum computing hardware, while remaining rooted into the physics-based modelling paradigm. Furthermore, its portability to off-lattice all-atom molecular representations paves the ways to perspective realistic applications.
Collectively, our results suggest that, if the size and performance of quantum simulators continue to improve over the next several years, the integration of quantum annealers and classical machine learning may represent a transformative new paradigm with broad implications in several areas of Life Sciences and Pharmacology.

\section{Models and Methods}
\begin{figure}[t!]
    \centering
    \includegraphics[width=\columnwidth]{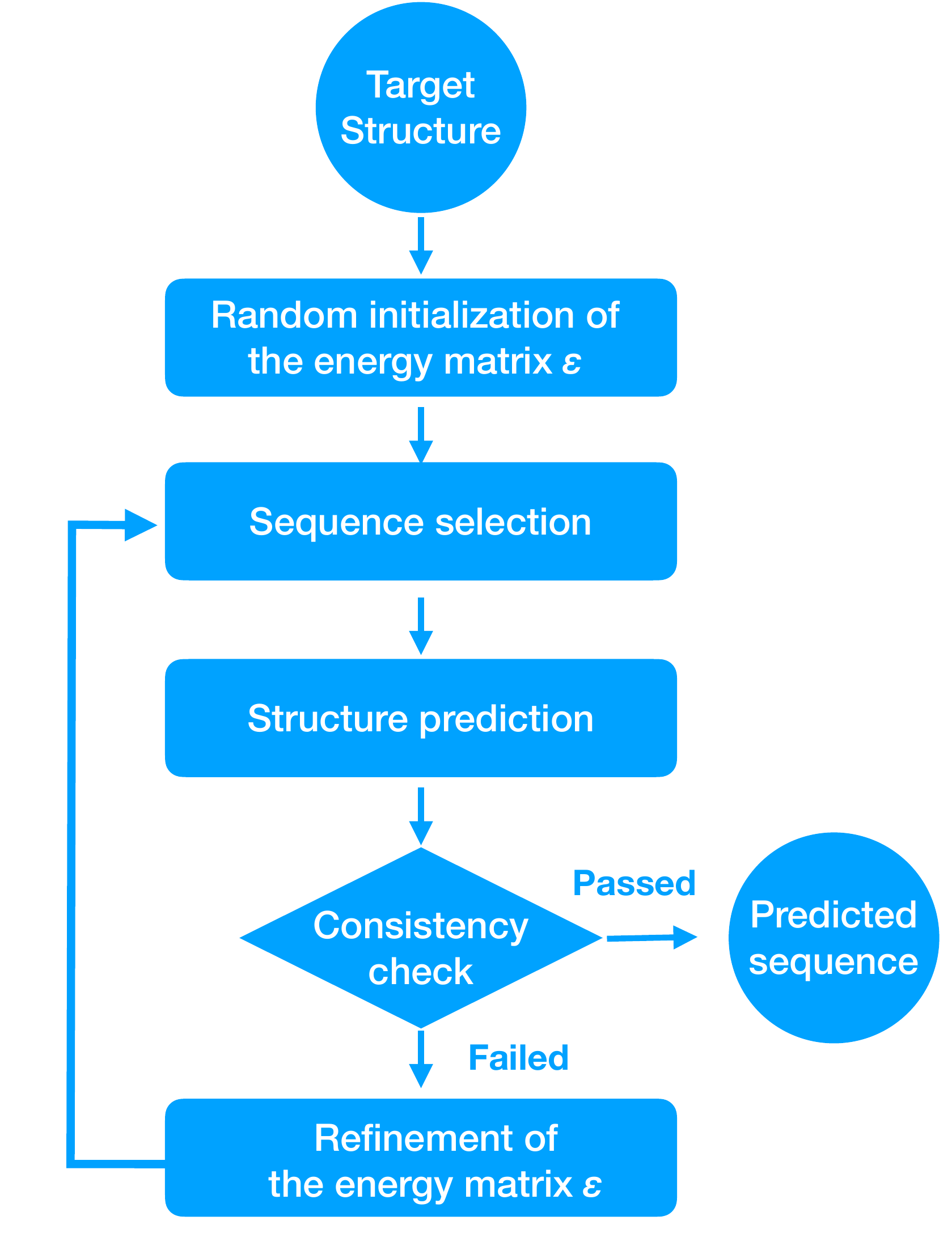}
    \caption{Schematic representation of our protein-design algorithm. Alternating steps of sequence selection and structure prediction are repeated. The energy map from an initial guess is updated until the algorithm passes consistency checks. The result is a sequence that folds into the desired target structure. Importantly, sequence selection and structure prediction can be integrated from different algorithmic or hardware platforms, permitting leveraging on rapid and complementary progress in technologies such as AI and quantum computing.}
    \label{fig:introductory_figure}
\end{figure}

\subsection*{Approximate scoring function $G(S)$}
We simplify the complexity of the design problem by introducing two approximations to circumvent the nested sequence-structure search implied by the minimization of the design scoring function in Eq.~\eqref{eq:exact_scoring_function}.

First,  we resort to a customary linear Ansatz for $E(\Gamma,S)$:
\begin{equation}
	E(\Gamma,S) \simeq \sum^\prime_{i,j=1...n} C_{ij}(\Gamma) \,\varepsilon_{s_i,s_j} \,.
	\label{eq:energy_functional}
\end{equation}
\noindent Here, $s_i$ is the chemical identity of the $i$th amino acid of the sequence $S$, which has length $n$, $\varepsilon_{s_i,s_j}$ are the entries of a suitable $D \times D$ energy matrix---$D$ being the size of the amino acids chemical alphabet---and $C(\Gamma)$ is the contact map of the conformational state $\Gamma$, with entries $C(\Gamma)_{ij}$ equal to 1 if amino acids $i$ and $j$ are in contact, and equal to 0 otherwise. The primed summation indicates the restriction to distinct pairs, $j>i$. A weighted contact map, with entries spanning the entire [0:1] interval, could also be used instead of the binary one.

In Eq.~\eqref{eq:exact_scoring_function}, the energy of the sequence $S$ mounted over the target structure is computed relative to the sequence free energy $F(S)$. Evaluating the latter implies computing the energy of $S$ mounted over all possible states, a computationally prohibitive task. A key approximation of our  approach  consists in replacing this reference with the average energy evaluated over a database of known native structures,
\begin{equation}
    F(S) \simeq \sum^\prime_{i,j=1...n} \varepsilon(s_i,s_j)\, \langle C_{ij} \rangle \ .
    \label{eqn:approxF}
\end{equation}
\noindent In the expression above, $\langle C_{ij} \rangle$ is the average contact map of those structures in the database that have the same length as the target one. In more general contexts, this restriction can be relaxed by setting $\langle C_{ij}\rangle$ equal to the contact probability of amino acids at chemical distance $|i-j|$ computed over all database entries that are sufficiently longer than $|i-j|$, to avoid end effects.

The gist of the approximation in Eq.~\eqref{eqn:approxF} is to yield a free energy estimate that, while remaining computationally amenable, is still informed by the structural properties of viable states. The average pairwise contact probabilities appear to be the most natural and effective choice in this respect, considering that the approach could be systematically generalized to include three-body and higher-order contact probabilities.

With this proviso, our approximation to the design scoring function $G(S)$ becomes
\begin{equation}
    G (S) \simeq \sum^\prime_{i,j=1...n} \varepsilon(s_i,s_j)\, (C_{ij}(\Gamma_T) - \langle C_{ij} \rangle) \ .
    \label{eq:approximated_scoring_function}
    \end{equation}
Minimizing this function thus selects sequences whose native energy is as low as possible compared to the average state in the database.
Importantly, the energy matrix $\varepsilon(s_i,s_j)$ does not need to be known \emph{a priori}. Instead, we propose an iterative approach, discussed in the next section, through which an initial guess is refined until consistency between the direct and inverse folding problems is reached.

\subsection*{Iterative scheme to tackle the design problem}
Key parameters of our design scheme are the entries of the $D\times D$ symmetric energy matrix $\boldsymbol{\varepsilon}$ of Eq.~\eqref{eq:approximated_scoring_function}.
Inspired by earlier work on the extraction of effective potentials for protein folding or design~\cite{maiorov1992contact,crippen1996easily,seno_micheletti_maritan_banavar_1998,Rossi_Micheletti_Seno_Maritan_2001}, we adopted an iterative scheme based on enforcing consistency between the solutions of the direct and inverse folding problems. Our choice is motivated by the increasing availability of reliable and fast algorithms for predicting protein folds even in realistic contexts~\cite{Jumper_Evans_Pritzel_et_2021}.
In principle, this opens the possibility of harnessing these efficient methods for tuning $\boldsymbol{\varepsilon}$ to design a specific type or family of target structures. Such optimized schemes would also limit adverse effects inherent to structural coarse-graining, which inevitably impacts the transferability of potential energies obtained by thermodynamic integration.
For the same reason, it is not apparent {\em a priori} that database-wide schemes for extracting interaction potentials, including the powerful quasichemical approximation ~\cite{Miyazawa_Jernigan_1985,skolnick1997derivation,tiana2004deriving,chen2005lessons,goldstein2007amino} for capturing amino acids interaction propensities, are ideally suited to the design task at hand based on the minimization of Eq.~\eqref{eq:approximated_scoring_function}.

The key steps of our iterative scheme are sketched in Fig.~\ref{fig:introductory_figure}. Given the target structure to be designed, $\Gamma_T$, and a random initialization of the energy matrix $\boldsymbol{\varepsilon}$, the scheme proceeds by iterating at each cycle the following steps:
\begin{enumerate}
\item \emph{Sequence selection}. Explore the combinatorial space of sequences to find the set $\mathcal{S} \equiv \{S_1, S_2, ...\}$ corresponding to the lowest values of $G(S)$.
\item \emph{Structure prediction.} For each sequence in $\mathcal{S}$ obtain a reliable prediction of the native state. We shall indicate such corresponding native set as $\mathcal{N}\equiv\{\tilde{\Gamma}_1, \tilde{\Gamma}_2,...\}$.
\item \emph{Energy function refinement}. Assess whether the states in $\mathcal{N}$ match the target structure $\Gamma_T$ within a specified tolerance. If so, the design problem of $\Gamma_T$ is solved, and the procedure ends. Otherwise, the symmetric energy matrix $\boldsymbol{\varepsilon}$ is refined to impose a consistency with the structure prediction results, i.e., to account for the fact that the native states of $\mathcal{S}$ do not include $\Gamma_T$.
\end{enumerate}

A fixed point in this iterative scheme embodies the highest achievable consistency between our heuristic scoring function and the chosen protein structure prediction algorithm.

{\bf Step 1: Sequence selection.} The first step of our iterative scheme involves solving a combinatorial optimization problem over the space of amino acid sequences. We will perform this step by constraining the overall amino acid composition, i.e., the abundances of the different types of amino acids. Thus, the first step involves minimizing Eq.~\eqref{eq:approximated_scoring_function} over the possible reshuffling of a given initial sequence with the desired composition. This task is an integer programming problem that can be carried out on conventional computers. However, considerable speed-ups for the same NP-complete class of problems may be achieved with quantum annealers, dedicated machines for solving quadratic unconstrained binary optimizations (QUBO) \cite{Hauke_Katzgraber_Lechner_Nishimori_Oliver_2020,Chang_Chen_Koerber_Humble_Ostrowski_2020,Jiang_Chu_2022,Au_Yeung_Chancellor_Halffmann_2023}.

To recast the minimization of $G(S)$ in Eq.~\eqref{eq:approximated_scoring_function} as a QUBO problem, we introduced binary variables to describe the chemical type, $s$, of each amino acid in the sequence. Specifically, to the $i$-th amino acid we associate an array of $D$ binary variables,  $\q{i}{m}$,  with $m = \{1,\dots,D\}$, with the proviso that, if $s_i$ is of type $j$, then $\q{i}{j}=1$ while the other $D-1$ array elements are equal to 0.

The variables are used to define a QUBO Hamiltonian whose ground states are in one-to-one correspondence with the sequences that minimize $G(S)$ at fixed composition.
This QUBO Hamiltonian consists of three terms, all quadratic in the $q$ entries,
\be
{\mathcal H} =  {\mathcal H}_{\rm comp} + {\mathcal H}_{\rm occ} + {\mathcal H}_{\rm contact}\ .
\label{eq:hamfull}
\ee
The first term attains its global minimum on sequences satisfying the composition constraint:
 \be
H_\mathrm{comp}=A_1\sum_{m=2}^{D}\pqty{\sum_{i\in\mathcal{L}}\q{i}{m} - N_m}^2\ ,
	\label{eq:contact_energy}
\ee
where $A_1>0$ and $N_m$ is the assigned number of amino acids of type $m$. Minimizing this term ensures that the non-zero entries of the $q$ arrays are consistent with the assigned composition.

The second term penalizes cases where more than one entry of the $\q{i}{m}$ array is equal to 1:
 \be
	H_\mathrm{occ}=A_2\sum_{i \in \mathcal{L}}\sum_{m\not=n=2}^{D} \q{i}{m}\q{i}{n}\ ,
 \ee
 with $A_2>0$. Minimizing this term along with $H_\mathrm{comp}$ ensures that the $q$ arrays encode a well-defined chemical species for each amino acid in $S$.

 The last term embodies the scoring function of Eq.~\eqref{eq:approximated_scoring_function}:
\be
    &&H_\mathrm{contact} =B\Big[\sum_{i,j}^{\prime}\sum_{m,n=2}^{D} \q{i}{m} \q{j}{n} C_{ij} \alpha_{mn}+ \nonumber \\
    &&+ 2\sum^{\prime}_{ij}\sum_{m = 2}^{D} \q{i}{m} C_{ij} \gamma_m + \,\sum^{\prime}_{i,j}C_{ij}\varepsilon_{11} \Big]\,. \label{eq:design_scoring_equation}
 \ee
\noindent In this equation, $B>0$ and $C = C\pqty{\Gamma_T} - \langle C \rangle$, where $C\pqty{\Gamma_T}$ is the adjacency matrix of the target configuration and $\langle C \rangle$ is the average contact map evaluated on a database of representative native structure. In addition, $\alpha_{mn} =\varepsilon_{mn} - \varepsilon_{m1} - \varepsilon_{n1} + \varepsilon_{11}$ and $\gamma_{m} = \varepsilon_{m1} - \varepsilon_{11} $.
The detailed derivation of this term is given in Section S1 of the SM. We note that the QUBO encoding of the minimization problem of Eq.~\eqref{eq:approximated_scoring_function}
is not unique; an example of an alternative QUBO encoding is presented in Section~S2 of the SM.

We emphasize that $H_{\textrm{occ}}$ and $H_{\textrm{comp}}$ encode the strong constraints, while $H_{\textrm{contact}}$ represents the molecular energy. As long as $A_1, A_2\gg B$, the ground state solutions of Eq.~\eqref{eq:hamfull}  simultaneously satisfy all the hard constraints and correspond to sequences that minimize $G(s)$ for the given chemical composition.

The sought ground states of the QUBO Hamiltonian can be found with various methods, some of which are compared here, including classical simulated annealing (see Section~S3 in the SM for details),  and the hybrid classical-quantum annealing scheme implemented in OCEAN, the user interface to the D-wave quantum annealer. The latter combines classical taboo search heuristic optimization  with quantum annealing  steps~\cite{Raymond_Stevanovic_Bernoudy_Boothby_McGeoch_Berkley_Farre_Pasvolsky_King_2023a}.

{\bf Step 2: Structure prediction.} The second step of the iterative scheme involves the application of structure prediction methods to the sequences identified from the minimization of $G(S)$ at the previous step. The key point is that the native states of such sequences are obtained with an independent  structure prediction method. In particular, the scheme used to predict native structures given a sequence is not informed by the energy matrix defining the design scoring function $G(S)$.

In realistic off-lattice applications, the go-to structure prediction methods would naturally be those based on heuristic machine-learning algorithms, which have proved to be reliable and efficient.
Considering the protein model's minimalistic nature, we opted for the most transparent and feasible method: an exhaustive search of conformational space to identify the lowest energy state(s) of a sequence based on a ground-truth energy matrix. This energy matrix is used solely in this step and for selecting viable target structures for the design problem, as detailed later, and is never used in the $G(S)$ definition.

{\bf Step 3: Energy Function Refinement.} The third step in the proposed iterative scheme involves updating the energy matrix entering $G(S)$ to improve consistency with the chosen ground-truth structure prediction algorithm.

To this end, we have devised the following scheme:
At the $k$-th step of the iterative procedure, let $S$ be a putative designing sequence obtained by minimizing $G$ based on the current energy matrix, $\boldsymbol{\varepsilon}^{(k)}$. The external protein structure prediction algorithm may find several structures for the sequence $S$ that are better faring than $\Gamma_T$ as native states. Let $\{\Gamma_0(S), \Gamma_1(S), \ldots,\Gamma_n(S)\}$ be a ranked set of such competing structures, ordered by increasing ground-truth energy, i.e., decreasing confidence score.

Since the structure prediction step is assumed to be reliable (and it certainly is in our minimalistic context where it entails an exhaustive search in structure space),  the observation that the competing structures $\Gamma_0,\dots, \Gamma_n$ have a higher confidence score than $\Gamma_T$ signals the imperfect parametrization of the $\boldsymbol{\varepsilon}^\pqty{k}$ matrix.

To compensate for this, we move to a new iteration where the energy matrix $\boldsymbol{\varepsilon}^\pqty{k+1}$ is updated over the $k$th one by requiring that $\Gamma_0,\dots, \Gamma_n$ have a lower energy than $\Gamma_T$, consistent with the outcome of the ground-truth predictor:
\be
\q[E]{k+1}{}(\Gamma_i,S) \leq \q[E]{k+1}{}(\Gamma_T,S), \qquad i \leq n,
\ee
\noindent where $E^{(k+1)}$ is the energy function of Eq.~\eqref{eq:energy_functional} informed by the interaction matrix $\boldsymbol{\varepsilon}^\pqty{k+1}$.
Let us consider the case where the best ranking structure $\Gamma_0$ has predicted probability above a given $p_\mathrm{fold}$.
In this case, we impose that the energy of $\Gamma_0$ should be significantly lower than that of all competing structures: $\forall i > 0$,
\be
    \q[E]{k+1}{}(S,\Gamma_i) \geq \q[E]{k+1}{}(S,\Gamma_0) +  \Delta\pqty{p_\mathrm{fold},\beta}\,,
    \label{eq:energy_refiniement_minimal_gap}
\ee
where $\Delta(p_\mathrm{fold},\beta) = \frac 1\beta \ln\frac{1-p_\mathrm{fold}}{p_\mathrm{fold}}$ is the minimum energy gap that would allow a sequence to fold into $\Gamma_0$ with probability $\geq p_\mathrm{fold}$ at inverse temperature $\beta$
(see Section~S3 of the SM for details).

Given that the coefficients to be learned enter linearly in the scoring function of Eq.~\eqref{eq:approximated_scoring_function}, fulfilling the set of inequalities is equivalent to solving a linear separability problem. This task can be conveniently tackled using established algorithms~\cite{krauth_mezard_1987,Kleinbaum_1994,Cortes_Vapnik_1995,Johnson_Mouhab_1996}.
In particular, in Section~S4 of the SM, we discuss our implementation based on the perceptron technique ~\cite{krauth_mezard_1987,Imhoff_1995, Freund_Schapire_1998,Du_Leung_Mow_Swamy_2022}, which has been previously used in different protein folding and design contexts to iteratively learn interaction potentials between amino acids~\cite{seno_micheletti_maritan_banavar_1998,dima_settanni_micheletti_banavar_maritan_2000}.

\subsection*{Lattice Protein Model}

We considered compact structures on a two-dimensional (square) lattice as a specific protein lattice model. The latter is customarily preferred over the cubic lattice because it offers a more realistic surface-to-volume ratio of compact structures of small length, $\lesssim 100$ amino acids. We consider sequence alphabets of $D= 3, 4$, and $5$ letters and target structures filling $4\times 4$, $5\times 5$ and $6\times 6$ lattices.

The $D \times D$ symmetric energy matrix, embodying the ground truth interaction potentials of the structure prediction step, was identified from a preliminary survey of viable combinations of interactions, meaning interactions that yield numerous designable structures. The latter correspond to structures that are the unique ground states of one or more sequences~\cite{li_emergence_1996}. The choice of ground truth energy matrix is detailed in Section~S6 of the SM.

\section{Results}
\label{sec:results}
In our iterative design strategy, the optimal parameters of the scoring function $G$ are obtained by comparing the results of direct vs. inverse folding predictions. Several factors may determine the quality of the predictions: (i) the feasibility of minimizing $G$ in the combinatorial space of sequences, (ii) the accuracy of the ``external" structure prediction method, (iii) the viability of the functional form of $G$ for yielding accurate design predictions when suitably parametrized, (iv) the feasibility of identifying such optimal parametrizations of $G$ using the iterative scheme.

In this proof-of-concept study, the uncertainties associated with points (ii) and (iii) are ruled out from the outset. Indeed, modeling proteins as compact structures on square lattices makes it possible to perform exhaustive searches in structure space, thus enabling the exact determination of the lowest energy state(s) of any given sequence. In addition, the functional form of the scoring function $G$, namely a pairwise-contacts Hamiltonian, was purposely chosen to match that of the ground-truth Hamiltonian used to pick designable structures as viable targets, thus guaranteeing that suitable parametrizations of $G$ exist and are, in principle, learnable.

In our context, where we shall use alphabets of limited size, $D=3$, $4$, $5$, point (i) could be addressed by exhaustive enumeration, similarly to the structure prediction step. However, in realistic contexts the exhaustive search of sequence or structure spaces would be unfeasible. While the structure space search can today be circumvented using the now available rapid and accurate structure prediction methods based on machine learning, the challenge of minimizing $G(S)$ in sequence space still persists. For this reason, we address point (i) by recasting the minimization of $G(S)$ as a QUBO problem, which can be tackled with classical and quantum combinatorial algorithms, see Methods. As we demonstrate by considering various values of $D$, such an approach is straightforwardly adapted to amino acid alphabets of any size.

\begin{figure}[H]
    \centering
    \subfloat[]{
        \includegraphics[width=0.3\linewidth]{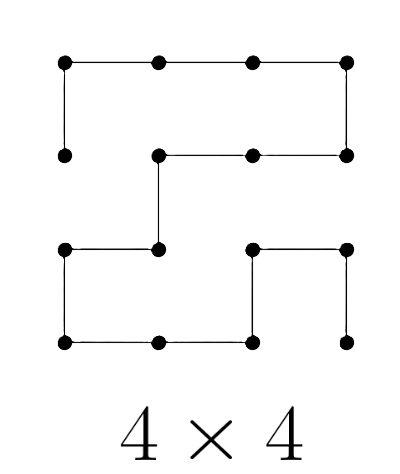}
        \label{fig:target_4}
        }\hfill
    \subfloat[]{
        \includegraphics[width=0.3\linewidth]{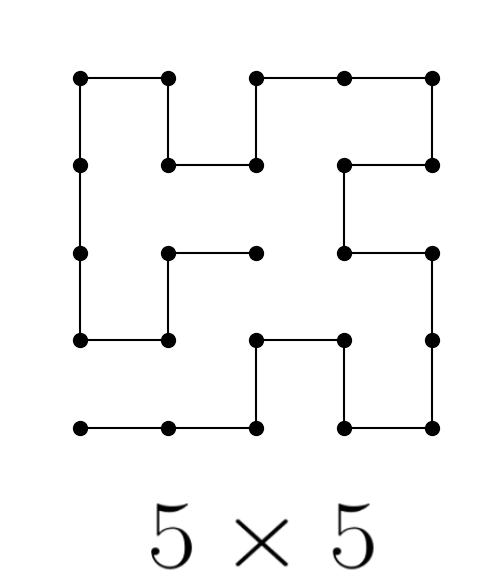}
        \label{fig:target_5}
    }\hfill
    \subfloat[]{
        \includegraphics[width=0.3\linewidth]{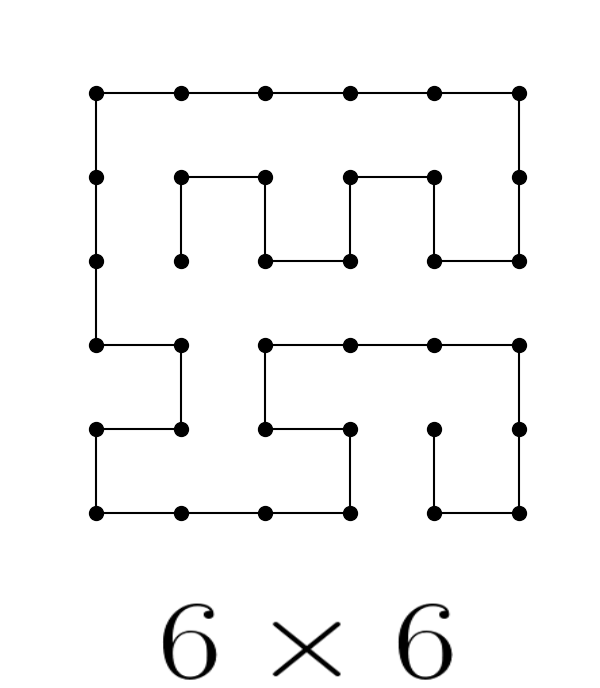}
        \label{fig:target_6}
    }
    \caption{Selected target structures for different lattice sizes.}
    \label{fig:targets}
\end{figure}

In connection to points (ii) and (iii) outlined above, we first assess the viability of the approximate functional form of  $G(S)$ as a scoring function to tackle the sequence optimization step. We considered two different contexts. In the first, the $\boldsymbol{\varepsilon}$ matrix is not learned but is set equal to that used in the ground truth protein folding predictor. In the second, $\boldsymbol{\varepsilon}$ is learned through our iterative procedure.

To carry out this assessment, we take the compact structure $\Gamma_T$ of Fig.~\ref{fig:targets}a as the design target. We use an alphabet of $D=3$ letters and set the composition to $\bqty{n_0=5,n_1=5,n_2=6}$, a choice that combines a sizeable combinatorial space of sequences with the existence of numerous solutions to the design problem.

For both $\boldsymbol{\varepsilon}$ choices, we computed $G(S)$ for all sequences with the above composition. We then obtained the receiver-operating curve (ROC), $y(x)$,  where $x$ is the rank index for increasing $G(S)$ and $y$ indicates what fraction of the exhaustive set of design solutions are found up to that value of the scoring function.

A perfect design performance would yield the steepest ramping ROC curve, where all the design solutions are exclusively at the highest-ranking positions (lowest values of the scoring function). Accordingly, a customary measure of ROC performance is the so-called normalized area under the curve, $Q$, i.e., the area between the curve and the diagonal divided by the area of the upper triangle. The aforementioned perfect performance would correspond to $Q \sim 1$. In contrast, in a baseline performance---where solutions are discovered with uniform probability independent of their $G(S)$ ranking---$Q$ would be close to $0$.

\begin{figure}
    \centering
    \includegraphics[width=0.8\linewidth]{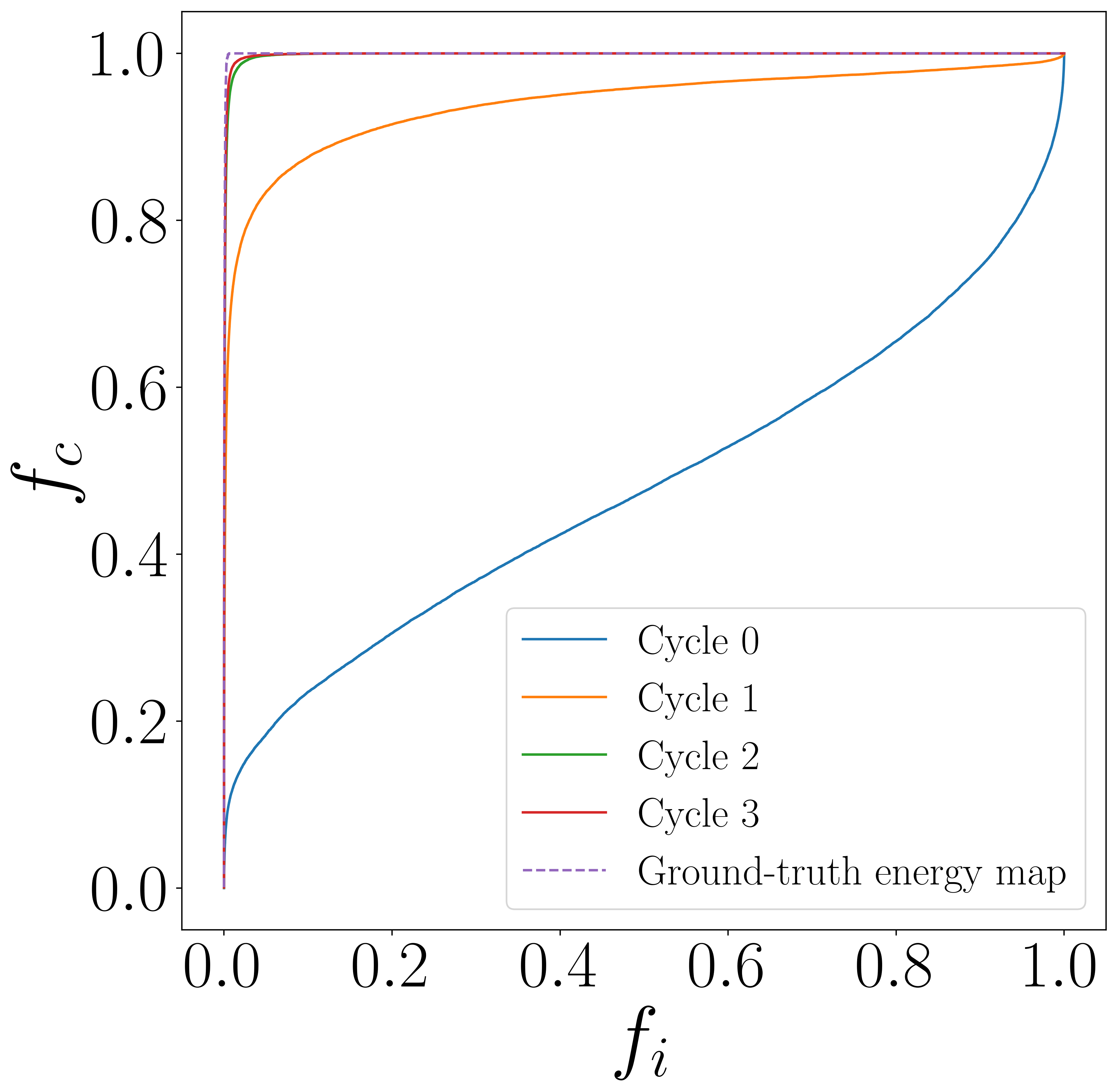}
    \caption{\textbf{Solid lines:} ROC curve at different iterations tests the goodness of $G(S)$ (see Eq.~\eqref{eq:approximated_scoring_function}) as a classifier, showing that it evolves from being nearly random (at cycle 0) to nearly optimal (at cycle 3). \textbf{Dashed line:} Plugging the ground-truth energy map in $G(S)$ produces a nearly-optimal classifier. }
    \label{fig:roc_curve}
\end{figure}

The results of our ROC analysis are shown in Fig.~\ref{fig:roc_curve}.
As anticipated, we preliminarily tested our $G(S)$ approximation by plugging the ground-truth potentials in place of the $\boldsymbol{\varepsilon}$ matrix.
The corresponding ROC curve, shown with a dashed line in Fig.~\ref{fig:roc_curve}, shows a near-ideal performance, $Q>0.99$. This demonstrates that the heuristic scoring function $G(S)$ of Eq.~\eqref{eq:approximated_scoring_function}, which is based on average contact probabilities, is indeed viable for design purposes, as it can lead to a near-perfect scoring when informed by suitable potentials.

We then moved to the second assessment, aimed at ascertaining if suitable parametrizations of $G(S)$ can be learned by our iterative design procedure starting from arbitrary initializations of the $\boldsymbol{\varepsilon}$ matrix. A further question is how many iterations are required for convergence.

For these tests, we applied the iterative procedure to the same designable target structure starting from 50 different random choices of the initial matrix.
The results are summarized in Fig.~\ref{fig:roc_curve}, which shows the ROC curves at different iteration stages, averaged over the different initializations---see Section~S7 of the SM for the individual ROC curves.

The blue curve in Fig.~\ref{fig:roc_curve} shows the average performance at the beginning of the iterative procedure (labeled cycle 0) when the energy matrix is yet to be learned. The curve is near-diagonal in this case, demonstrating the expected baseline performance.
The performance steadily improves at each iteration,  converging to a nearly perfect parametrization, corresponding to $Q>0.99$, in as few as three iterations.

\paragraph{Performance scaling with lattice and alphabet size.}
\label{par:scaling_alphabet_lattice}

\noindent We next turned to larger lattice sizes and amino acids alphabets, see SM. In such cases, ROC curves are not the best way to assess the design performance as they require exhaustive coverage of sequence space, which becomes rapidly impractical with growing protein length and alphabet size.

Instead, we estimated the design success rate using a sampling scheme. Specifically, at each iteration, we selected the 30 best-scoring equences according to $G(S)$ and computed which fraction of them, $f_c$, admitted the target structure as the unique ground state and satisfied Eq.~\eqref{eqn:des2}.

\begin{figure}
\subfloat[]{
    \includegraphics[width=0.9\linewidth]{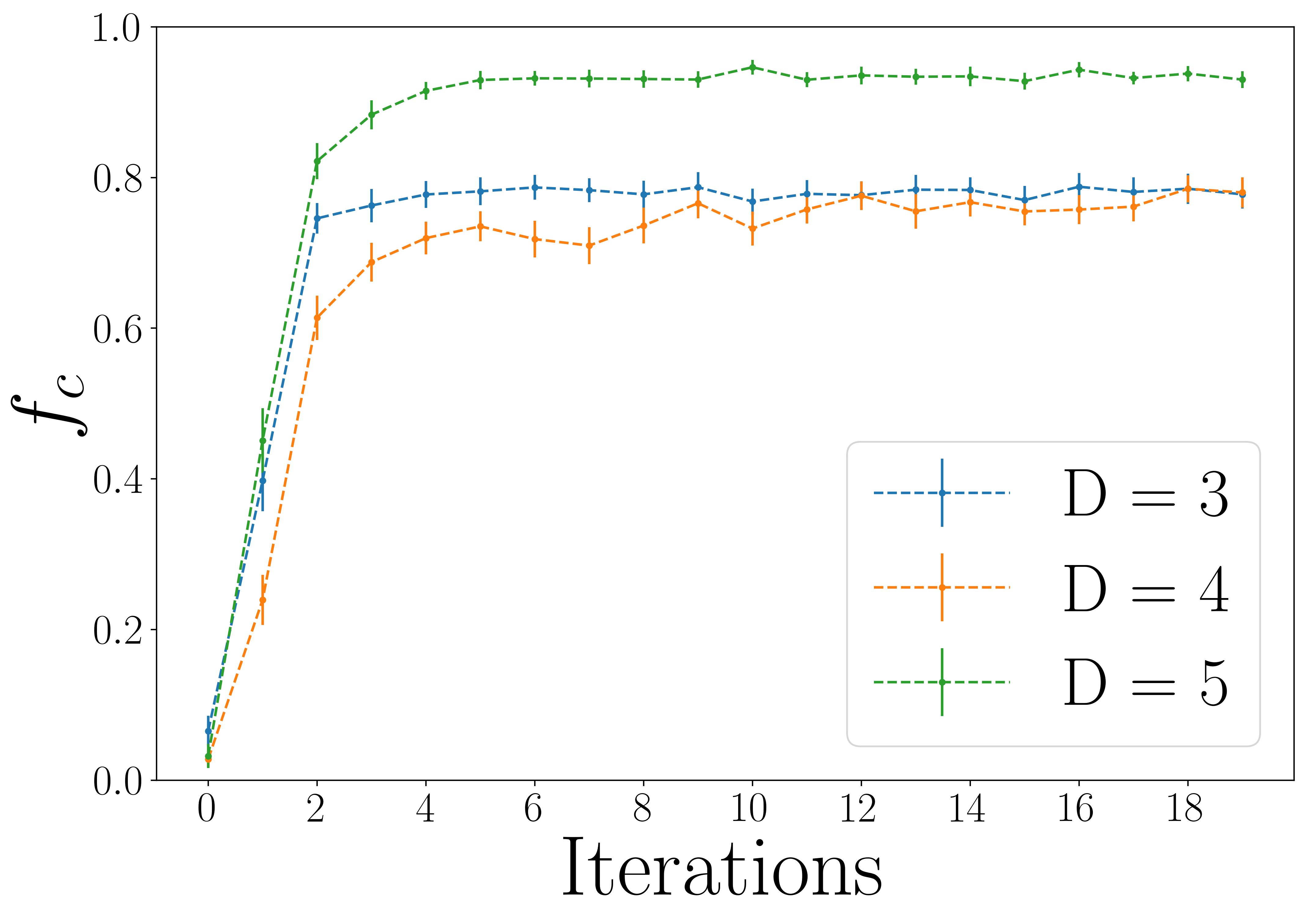}\label{fig:alphabet_scaling}
    } \hfill
\subfloat[]{
    \includegraphics[width=0.9\linewidth]{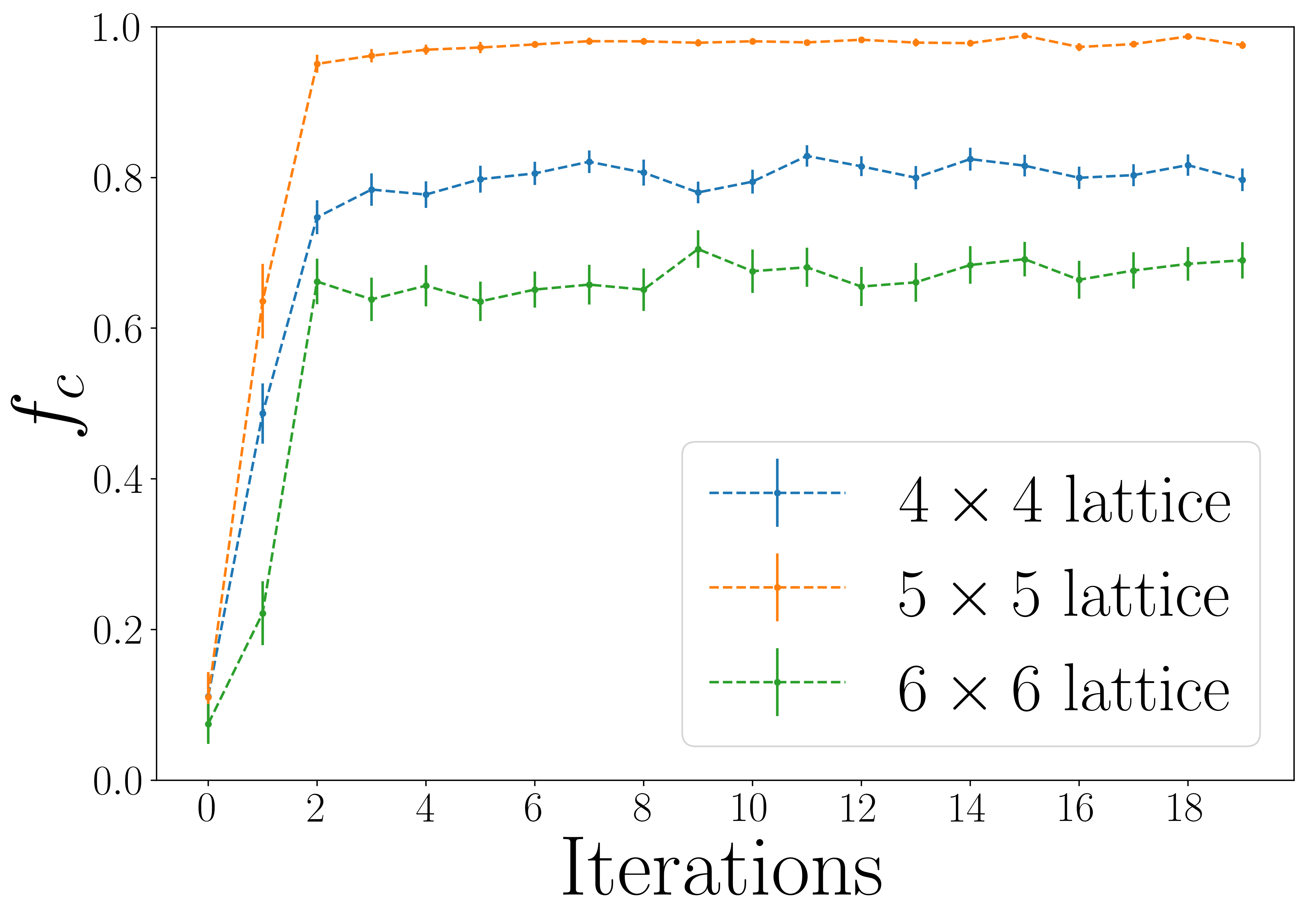}\label{fig:lattice_scaling }
    }
    \centering
    \caption{Fraction $f_c$ of correctly identified sequences as a function of refinement iterations for different alphabet and lattice sizes. In \textbf{(a)}, we consider a fixed structure on a $4 \times 4$ lattice, see Fig.~\ref{fig:target_4}, and vary the size $D$ of the alphabet. In \textbf{(b)}, we fix the alphabet size $D=3$ and span over different lattice sizes. The corresponding target structures are in Figs.~\ref{fig:target_4},~\ref{fig:target_5}, and \ref{fig:target_6}.}
    \label{fig:scalability}
\end{figure}

The results are given in Fig.~\ref{fig:alphabet_scaling} and show that, for all three alphabet sizes considered, $D=\{3,4,5\}$, our algorithm reaches a success rate of about $80\%$ after just a few iterations. Notably, the highest performance is achieved with the largest alphabet size, corresponding to 5 amino-amino acid types. Importantly, this trend is robust over different choices of the target structure (see Section~S8 of the SM).

In Fig.~\ref{fig:lattice_scaling }, we report the results of a similar analysis for a fixed alphabet of $D=3$ letters but for three compact structures filling lattices of increasing sizes.
Again, in all cases, the algorithm reaches a plateau after a few iterations. While these specific instances do not show a clearly identifiable trend with lattice size, when the analysis is extended to an ensemble of structures, we observe that the success rate decreases with increasing chain size (see Section S8 of the SM).

For the systems we have considered, the overall success rate ranges from about  $65\%$ up to nearly 100\%.

\paragraph{Comparison of conventional and QUBO-based minimizers.}
\label{par:comparison_classical_quantum}

A key feature of our approach, is that the combinatorial search underpinning the sequence selection step is formulated as a QUBO problem, which is, in principle, amenable to quantum annealers.
This poses two questions: (i) Does the QUBO encoding boost the design performance compared to working directly in sequence space? (ii) How do currently available quantum annealers fare at the design task compared to state-of-the-art classical QUBO solvers?

\begin{figure}
\centering
    \subfloat[]{\includegraphics[width=\linewidth]{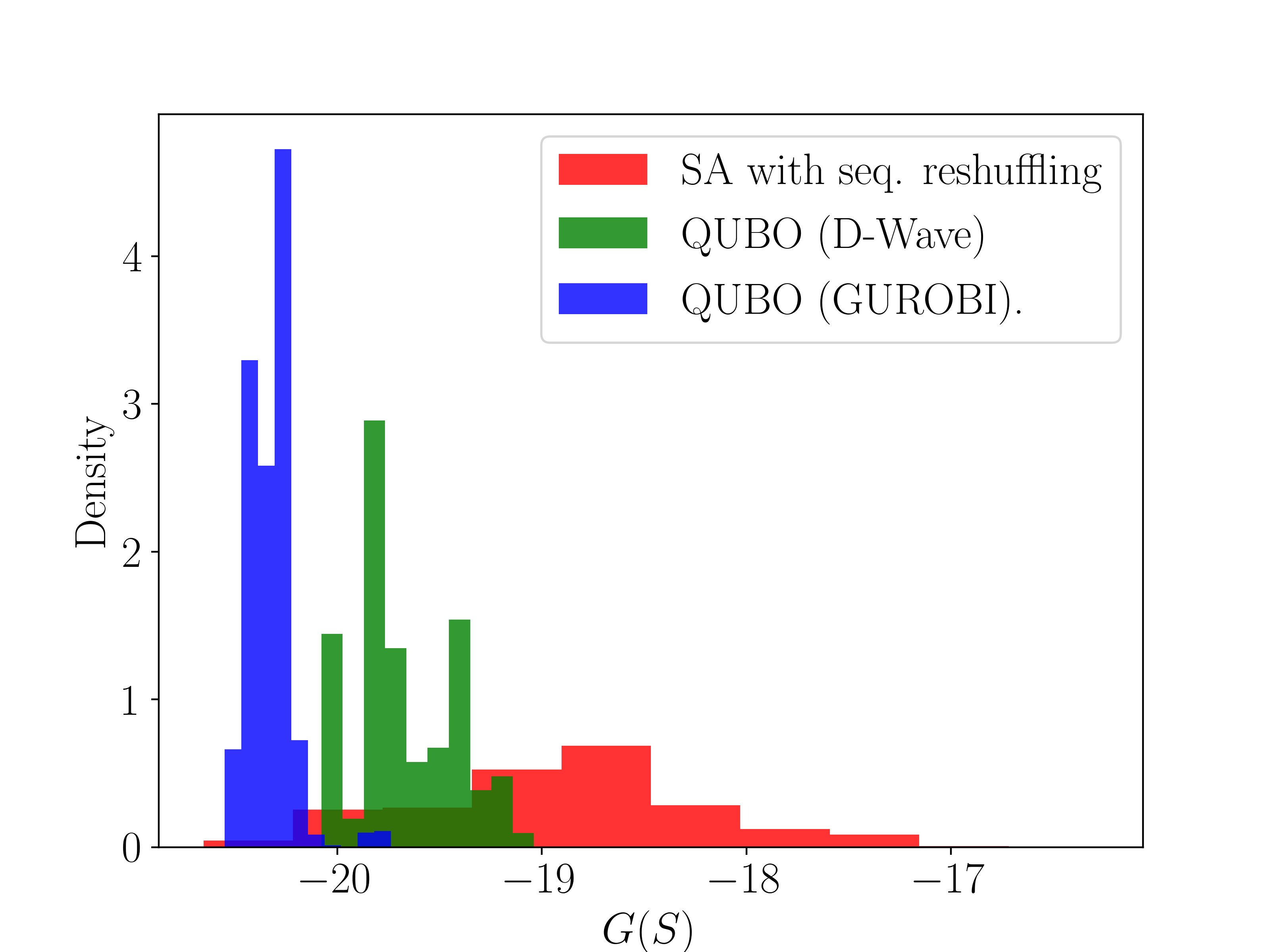}}\\
    \subfloat[]{\includegraphics[width=\linewidth]{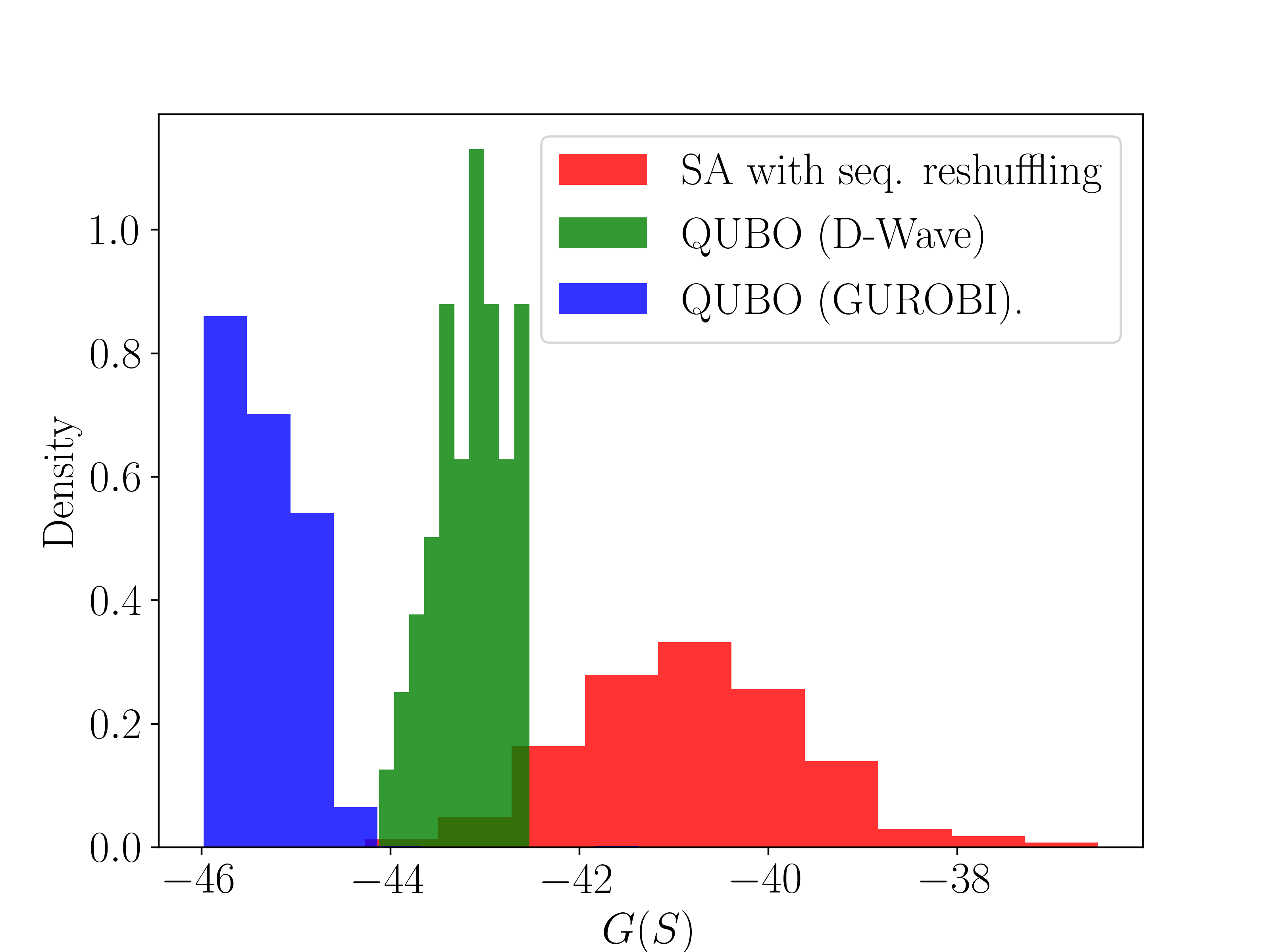}}
    \caption{Statistics of the $G(S)$ values resulting from the use of different optimization approaches (simulated annealing, hybrid annealing on D-WAVE, and GUROBI optimizer).
    In particular, we represent 1000 samples obtained by using simulated annealing, 1000 using the GUROBI optimizer, and 100 using the D-Wave hybrid optimizer.
    In \textbf{(a)} we consider a structure on a $9 \times 9$ lattice, while in \textbf{(b)} we consider a $13 \times 13$ lattice.}
    \label{fig:histograms}
\end{figure}

To address these questions, in Fig.~\ref{fig:histograms}, we report, for compact structures of different sizes, the lowest values of $G(S)$ obtained after many runs of classical optimizations of the scoring functions, parametrized with the ground-truth potentials, with different encodings and hardware at equal duration (3s). The red histogram corresponds to the results of simulated annealing directly formulated in sequence space, where the moves correspond to composition-preserving reshufflings of the sequence. Instead, the blue curve was obtained using GUROBI, an industry-grade QUBO solver. Finally, the green histogram reports the results of 3s runs on D-Wave using the hybrid classical-quantum solver, given that the complexity of the problem at hand exceeds the size currently addressable with fully-quantum annealing algorithms.

The most striking feature of these results is that the $G(S)$ distributions of the minimizers based on the QUBO formulation (green and blue) extend well below the lower tail of the distribution generated with a conventional optimization based on the combination of sequence reshuffling and simulated annealing (red).
Hence, the QUBO reformulation required to harness quantum computing technologies has generated a major improvement in the sequence optimization step even when adopted on classical machines.  Notably, this difference in performance is enhanced for the largest lattice size.

Focusing on QUBO solvers, we note that the best performance is achieved by GUROBI, an entirely classical scheme based on heuristic searches. This result highlights the maturity reached by classical optimizers following from decades of hardware and software development. At the same time, we emphasize that the hybrid scheme implemented in D-Wave interleaves classical and quantum steps with internal criteria that are not easily controllable by the user. Thus, the results of the hybrid algorithm arguably represent a lower bound of the performance achievable by optimal combinations of classical and quantum steps.

\section{Conclusion and Perspectives}
\label{sec:discussion_conclusions}

In this proof-of-concept study, we have shown that the availability of reliable algorithms for protein structure predictions can be capitalized to envision efficient strategies for tackling the protein design problem. There are two distinctive features of our iterative method. First, the structure prediction algorithms are used to learn an optimal scoring function for the design problem instead of using postulated models and interaction parameters. Secondly, having mapped the sequence selection step to a combinatorial QUBO problem allows for addressing the design problem by harnessing existing powerful classical optimizers and promising quantum technologies.

Strikingly, we found that the QUBO encoding brings {\em per se} a significant improvement, to the point that matching the performance of classical or quantum QUBO optimizers with conventional schemes becomes computationally impractical even for modest protein lengths.

In our first illustrative benchmarks, we resorted to exhaustive enumeration to remove the uncertainty associated with heuristic protein folding predictors. This choice had the downside of considering simplified lattice models and relatively small chains and alphabets.
However, we note that the key theoretical ingredient of our method is the scoring function $G(S)$ that is entirely specified in terms of contact maps, regardless of the specific representation of protein conformations. As such, our method is primed to be generalized to off-lattice contexts.
By moving to realistic representations, one could capitalize on state-of-the-art machine-learning predictors \cite{Jumper_Evans_Pritzel_et_2021}, which typically return accurate folding solutions in seconds, a time much shorter than what is required by our exhaustive enumeration on lattice models.

Furthermore, the trajectory in addressing the technological limitations of the existing quantum hardware has been impressive~\cite{Pochart_Jacquot_Mikael_2022,Hauke_Katzgraber_Lechner_Nishimori_Oliver_2020}. This gives hope that significantly more complex problems will become solvable in the near future, opening the possibility of optimizing $G(S)$ in the realistic sequence alphabet space.

Succeeding at this task will be transformative not only for protein design. However, it would also pave the way towards a broad range of related applications, e.g., protein origami, drug screening, and {\em de novo} drug design.

\subsection*{Acknowledgments}

We are grateful to Francesco Slongo for his help with numerical work and to Davide Pastorello and Enrico Blanzieri for useful discussions.

This work was supported partly by Qub-IT, a project funded by the Italian Institute of Nuclear Physics (INFN) within the Technological and Interdisciplinary Research Commission (CSN5), partly by PNRR MUR projects under Grants PE0000023-NQSTI and CN00000013-ICSC, partly by PNRR grant CN\_00000013\_CN-HPC, M4C2I1.4, spoke 7, funded by NextGenerationEU, and partly by MUR grant PRIN 2022R8YXMR.

PH and VP acknowledge support from Q@TN, the joint lab of the University of Trento, FBK-Fondazione Bruno Kessler, INFN-National Institute for Nuclear Physics, and CNR-National Research Council.

PH has further received funding from the European Union’s Horizon Europe research and innovation programme under grant agreement No 101080086 NeQST and the Italian Ministry of University and Research (MUR) through the FARE grant for the project DAVNE (Grant R20PEX7Y3A).
This project was funded by the European Union under NextGenerationEU via the ICSC – Centro Nazionale di Ricerca in HPC, Big Data and Quantum Computing,
PH also acknowledges support from Provincia Autonoma di Trento and from ICSC – Centro Nazionale di Ricerca in HPC, Big Data and Quantum Computing, funded by the European Union under NextGenerationEU.
Views and opinions expressed are however those of the author(s) only and do not necessarily reflect those of the European Union, The European Research Executive Agency, or the European Commission.  Neither the European Union nor the granting authority can be held responsible for them.

PF  acknowledges support from by U. Milan-Bicocca's Center for Quantum Technologies (BiQuTe).

\subsection*{Conflict of Interests}
PF is a co-founder and shareholder of Sibylla Biotech SPA, a company developing and employing advanced molecular simulations for early-stage drug discovery.

\begin{titlepage}
  \centering
  \vskip 1em
  \large \textbf{Protein Design by Integrating Machine Learning with Quantum Annealing and Quantum-inspired Optimization.
Supplementary Material }\vskip 1em
  \small Veronica Panizza$^{1,2}$, Philipp Hauke$^{1,2}$, Cristian Micheletti$^3$, Pietro Faccioli$^4$\\\vskip 0.2em
  \small\textit{
  $^1$Pitaevskii BEC Center,  Physics Department, Trento University, Via Sommarive 14, 38123 Povo (Trento), Italy.\\
  $^2$INFN-TIFPA, Via Sommarive 14, 38123 Povo (Trento), Italy.\\
  $^3$Scuola Internazionale Superiore di Studi Avanzati (SISSA),Via Bonomea 265, I-34136 Trieste, Italy.\\
  $^4$Department of Physics University of Milano-Bicocca and INFN, Piazza della Scienza 3, I-20126 Milan, Italy.}
  \vskip 2em

\end{titlepage}
%

In this Supplementary Material, we provide details concerning implementing the protein design algorithm described in the main text and its first illustrative application. 
Namely, in Section S1, we illustrate how the design scoring function (Eq.(6) of the main text) is recast as a QUBO functional, while in section S2 we propose an alternative encoding.
Section S3 details the simulated annealing scheme used in our work. 
In Section S4, we derive the term $\Delta(p_\mathrm{fold},\beta)$, entering in Eq.~(12) of the main text.
In Section S5, we describe our application of the perceptron technique \cite{krauth_mezard_1987} to refine the entries of the energy matrix entering the design scoring function in Eq.~(6) of the main text.   
In Section S6,  we report all parameters required to reproduce the results of the illustrative application described in the main text. In Section S7, we include the individual Receiver Operation Curves (ROC) used to compute the averages ROCs reported in the main text. 
Section S8 discusses the impact of increasing alphabet or lattice size on the overall success rate of our algorithm.

\section*{S1: QUBO Encoding of the  Design Scoring Function}
\label{sec:design_scoring_function}

In this section, we explicitly derive the QUBO Hamiltonian that encodes the combinatorial optimization problem associated with minimizing the design scoring function $G(S)$ defined in Eq.~(6) of the main text.

To set the stage, we consider a lattice $\mathcal{L}$ composed of $N$ beads, a D-dimensional alphabet $\mathcal{D} = \Bqty{1,2,\dots,D}$, 
and a composition array $\vb{N}=\Bqty{N_1, N_2, \dots,N_{D}}$, with $\norm{\vb{N}}_1= N$. The composition array specifies the number of residues of each type in the chain.  

If $\q{i}{m}$ is 1 (0) then monomer of type $m$ is (not) present at site $i$.
To prevent two or more monomers from occupying in the same site and to impose the global chemical composition given by the occupation array $\vb{N}$, we enforce
\begin{equation}
	\forall \, i \in \mathcal{L} \quad \sum_{m\in \mathcal{D}}\q{i}{m} = 1 \qand
	\forall\, m\in \mathcal{D} \quad \sum_{i\in\mathcal{L}} \q{i}{m} = N_m\,.
	\label{eq:dependent_constraints}
\end{equation}
The constraints in Eq.~\eqref{eq:dependent_constraints} are not linearly independent;
to see this, let us consider $\sum_{m\in\mathcal{D}}\sum_{i\in\mathcal{L}}\q{i}{m}$ and, 
using the composition constraint, express it as $\sum_{m\in\mathcal{D}} N_m = N$. 
The same result can be derived using the excluded volume condition; 
then, we reduce constraints in Eq.~\eqref{eq:dependent_constraints} to
\begin{equation}
	\forall\, m \in \mathcal{D}_1 \quad \sum_{i\in\mathcal{L}}          \q{i}{m} = N_m\qand
	\forall\, i \in \mathcal{L}       \quad \sum_{m \in\mathcal{D}_1} \q{i}{m} \leq 1\,,
	\label{eq:independent_constraints}
\end{equation}
where $	\mathcal{D}_1 = \{2, \dots, D\}$ is a reduced alphabet. 
In this setting, $\q{i}{1}$ is bound to be $1-\sum_{m\in \mathcal{D}_1}\q{i}{m}$. With this, we conclude that $N(D-1)$ qubits are sufficient to enforce all constraints in our simulation. 
To embed these constraints in our simulation, we need to express them in terms of quadratic penalties. 
Since $\q{i}{m}\in \{0,1\}$, the excluded volume condition is satisfied iff $\sum_{m\in\mathcal{D}_1}\q{i}{m}$ is 0 or 1. 
This is equivalent to enforce $\pqty{\sum_{m\in\mathcal{D}_1}\q{i}{m}}^2 = \sum_{m\in\mathcal{D}_1}\q{i}{m}$, 
and leads to the quadratic penalty term 
\be
\sum_{i\in\mathcal{L}}\sum_{m\not =n\in \mathcal{D}_1} \q{i}{m}\q{i}{n}\,.
\ee 
Composition-violating configurations are penalized by 
\be
\sum_{m\in\mathcal{D}_1}\pqty{\sum_{i\in\mathcal{L}} \q{i}{m} - N_m }^2\,.
\ee

 Let us now show that Eq.~(10) of the main text follows from Eq.~(6) of the main text, which we recall being 
\be
G(\Gamma,S) = \sum_{i,j}^\prime C_{ij}\, \varepsilon_{s_i,s_j}\,,
\ee
where $C = C(\Gamma) - \langle C \rangle$. With our definition of $\q{i}{m}$, the quantity $\varepsilon_{s_i,s_j}$ is $\sum_{m,n \in \mathcal{D}} \q{i}{n}\q{j}{m} \varepsilon_{nm}$, so that
\begin{align}
	G(\Gamma,S) = &\sum_{i,j}^\prime\sum_{m,n \in \mathcal{D}_1} C_{ij}\, \q{i}{m} \q{j}{n} \, \varepsilon_{mn}+ \nonumber\\
	&\sum_{i,j}^\prime\sum_{m \in \mathcal{D}_1}    C_{ij}\, \q{i}{m} \q{j}{1} \, \varepsilon_{m1}+\nonumber \\
	&\sum_{i,j}^\prime\sum_{n \in \mathcal{D}_1}     C_{ij}\, \q{i}{1} \q{j}{n} \, \varepsilon_{1n}+ \nonumber\\ 
	&\sum_{i,j}^\prime C_{ij}\, \q{i}{1} \q{j}{1} \, \varepsilon_{11}\,.
\end{align}
Using the $\sum_{m\in \mathcal{D}} \q{i}{m} = 1$ condition yields
\begin{align}
	G(\Gamma,S) = &\sum_{i,j}^\prime\sum_{m,n \in \mathcal{D}_1} C_{ij}\, \q{i}{m} \q{j}{n} \, \varepsilon_{mn}+ \\
	&\sum_{i,j}^\prime\sum_{m \in \mathcal{D}_1}    C_{ij}\, \q{i}{m} \pqty{1-\sum_{n\in \mathcal{D}_1}\q{j}{n}}\, \varepsilon_{m1}+\nonumber \\
	&\sum_{i,j}^\prime\sum_{n \in \mathcal{D}_1}     C_{ij}\, \pqty{1-\sum_{m\in \mathcal{D}_1}\q{i}{m}} \q{j}{n} \, \varepsilon_{1n} + \nonumber\\ 
	&\sum_{i,j}^{\prime} C_{ij}\, \pqty{1-\sum_{n\in\mathcal{D}_1}\q{i}{n}}\pqty{1-\sum_{m\in\mathcal{D}_1} \q{j}{m}} \, \varepsilon_{11}\,, \nonumber
\end{align}
and finally
\begin{align}
	G(\Gamma,S) = &\sum_{i,j}^{\prime}\sum_{m,n \in \mathcal{D}_1}C_{ij}\,\q{i}{m} \q{j}{n} \pqty{\varepsilon_{mn} - \varepsilon_{m1} - \varepsilon_{1n} + \varepsilon_{11}} + \nonumber\\
	&+2\sum_{i,j}^{\prime}\sum_{m \in \mathcal{D}_1}C_{ij}\,\q{i}{m} \pqty{\varepsilon_{1m} - \varepsilon_{11}}+\nonumber\\ 
        &+\sum_{i,j}^{\prime} C_{ij}\, \varepsilon_{11}\,.
\end{align}

\section*{S2: Alternative Logarithmic Encoding}
\label{sec:alternative_encoding}
In this section, we provide a logarithmic encoding of Eq.~(6) of the main text as an alternative to Eq.~(7) through~(10) of the main text.
For the sake of simplicity let us consider a $D = 2^d$ -- dimensional alphabet, allowing us to describe the state of a vertex $i$ using an array of $d$ qubits (i.e. $\q[\vb{q}]{i}{} = (\q{i}{d-1},\dots, \q{i}{1},\q{i}{0})$). In this setting, if the $n$--th monomer sits in $i$, then $\q[\vb{q}]{i}{} = \pqty{n_{d-1},\dots,n_1,n_0}$, with $n_{d-1}\dots n_1 n_0$ being the binary representation of $n$.
To embed Eq.~(6) of the main text using this encoding requires us to introduce a function $\delta(n,\q[\vb{q}]{i}{})$ that is 1 iff $q^{(i)}$ accounts for the presence of monomer $n$ in site $i$. 
With this,
\be
    G(\Gamma,S) = \sum_{i,j}^\prime\sum_{n,m} C_{ij}(\Gamma)  \, \delta(n,\q[\vb{q}]{i}{}) \, \delta(m,\q[\vb{q}]{j}{}) \varepsilon_{nm}\,.
    \label{eq:alternative_encoding}
\ee
In general, to $\delta(n,\q[\vb{q}]{i}{})$ contribute 1, 2, \dots, d--body terms; then, to write Eq.~\eqref{eq:alternative_encoding} as a quadratic expression, every $\delta(n,\q[\vb{q}]{i}{})$ needs to be recast as a single-body term by adding ancillary qubits. 
In particular, each n--body term results from ``gluing" two $(n-1)$--body terms using an acillary qubit (e.g., $(q_1q_0) =(q_1)(q_0)$,\,$(q_2q_1q_0) = (q_2q_1)(q_1q_0),\, (q_3q_2q_1q_0)=(q_3q_2q_1)(q_2q_1q_0)$, etc.). 
Since the number of $n$--body terms that can be enstablished among $d$ qubits equals $\frac{d!}{n! \, (d-n)!}$, the total number of qubits required to ancillarize such terms amounts to $\frac{d!}{n! \, (d-n)!}$.
Full rewriting of $\delta(n,\vb{q})$ as single-body term requires us to take care of 2, 3, \dots, $d$--body terms, so that the total amount of ancillas employed for each vertex is
\begin{equation}
	\sum_{n=2}^{d} \frac{d!}{n!\,(d-n)!} = \sum_{n=0}^{d} \frac{d!}{n! \,(d-n)!} - 1 - d = 2^d - 1 - d\,.
\end{equation}
\noindent Summing the $d$ qubits per vertex used to encode the sequence, we need a total of $2^d - 1 = D - 1$ qubits per vertex. 
Compared to the encoding in S1, there is no saving in qubit resources used. 
In the main text and in our numerics, we have opted for using the encoding in S1 as it is physically more transparent.  

\section*{S3: Simulated annealing on classical machine}
\label{sec:simulated_annealing}
In this section, we provide information regarding the  classical combinatorial optimization algorithms employed in  the sequence optimization step of our iterative algorithm.
In our approach, we employed an improved version of the simulated annealing scheme to retain only trial moves that preserve the global relative abundance of chemical identities in the chain, thereby ensuring a much larger acceptance rate. In this approach, the chain sequence was not specified by means of binary variable (as in the QUBO formulation), but rather directly as a sequence $(X_1,X_2,\dots)$ of alphabet elements ( $X_i \in \mathcal{D}$).
Trial moves were proposed by swapping randomly chosen  pairs of monomers in the sequence. 
For example, given a sequence  $S= (X_1,X_2,\dots)$  of alphabet elements, selecting $i$-th and $j$-th monomers, proposes the sequence
\be
S' = (\dots, X'_i,\dots,X'_j,\dots) = (\dots,X_j,\dots,X_i,\dots)\,.
\ee
We stress that this prescription automatically avoids  configurations where several monomers occupy the same site; in addition to this, since moves conserve the composition of a given sequence, it is sufficient to select as initial state a sequence that meets the composition constraint.

In all annealing simulations, we relied on the \texttt{Simanneal} package for python~\cite{simanneal}, with the following choice of parameters: $T_\mathrm{max} = 100$, $ T_\mathrm{min} = 10^{-4}$, and $N_\mathrm{steps} = 10^4$.

\section*{S4: Minimal energy gap}
\label{sec:minimal_energy_gap}
In this section, we derive the term $\Delta(p_\mathrm{fold},\beta)$ contributing to Eq.~(12) of the main text.
This is the minimal energy gap separating ground-- and first--excited state of a protein-like sequence $S$, i.e., satisfying
$\frac{-\beta E(S,\Gamma_0)}{\sum_{\Gamma'} e^{-\beta E(S,\Gamma')}}\geq p_\mathrm{fold}$. 
Considering the ordered spectrum of $S$, $\gamma = \Bqty{\Gamma_0,\Gamma_1, \dots, \Gamma_M}$, we rewrite this expression as
\begin{equation}
	1 \geq p_\mathrm{fold} \pqty{ 1 + e^{-\beta \Delta_{01}} + e^{-\beta \Delta_{02}} + \dots + e^{-\beta \Delta_{0M}}}\,, 
\end{equation}
where $\Delta_{0n} = E(S,\Gamma_n) - E(S,\Gamma_0)$. 
If the previous inequality holds, then
\begin{align}
	\frac 1p_\mathrm{fold}  - 1 &\geq e^{-\beta \Delta_{01}} \nonumber\\ 
        &\Updownarrow\nonumber\\
        \Delta_{01} \geq \frac 1\beta \ln \frac{1-p_\mathrm{fold}}{p_\mathrm{fold}}& = \Delta(p_\mathrm{fold},\beta)\,,
\end{align}
setting a necessary but not sufficient condition for $S$ to be protein-like (i.e. fold to native state with a probability higher than $p_\mathrm{fold}$) at inverse temperature $\beta$. As a consequence, the constraint in which $\Delta(p_\mathrm{fold},\beta)$ is involved is quite a loose constraint; nonetheless, it can be employed without requiring to know the details of the ground-truth energy spectrum of the considered sequence.

\section*{S5: Solving our classification problem with the perceptron technique}
\label{sec:perceptron_technique}
In our iterative protein design scheme, the energy matrix entering the heuristic scoring functional in Eq.~(6) of the main text is iteratively learned by comparing the results of direct and inverse folding predictions. 
As discussed in the main text, this step involves satisfying the conditions in Eqs.~(11) and~(12) of the main text, a task which can be solved using different methods \cite{krauth_mezard_1987,Kleinbaum_1994,Imhoff_1995,Cortes_Vapnik_1995, Johnson_Mouhab_1996,Freund_Schapire_1998,Du_Leung_Mow_Swamy_2022}.\\

In our designing algorithm, $\boldsymbol{\varepsilon}^\pqty{k+1}$ is required to satisfy a heuristic set of conditions that depend on the set $\mathcal{S}^{(k)}_\mathrm{cum.}$ formed by all the sequences that have been generated by minimizing $G(S)$ up to $k$--th iteration:  $\mathcal{S}^{(k)}_\mathrm{cum.} = \mathcal{S}^{(k)}\cup \mathcal{S}^{(k-1)}\cup\dots \cup \mathcal{S}^{(0)}$.\\ 

We adopt the notation $\Gamma_i \prec_S \Gamma_T$ if the ground-truth energy of  sequence $S$ mounted on $\Gamma_i$ is lower than the one of $S$ mounted on $\Gamma_T$.
With this, the set of constraints is expressed as
\begin{gather}
    \underbrace{\forall \textrm{ S} \in \mathcal{S}^\pqty{k}_\mathrm{cum.}\qc \Gamma_i \preceq_S \Gamma_T}_{\Downarrow} \nonumber \\
    E^{(k+1)}(\Gamma_i,S) \leq E^{(k+1)}(\Gamma_T,S),\nonumber    
\end{gather}
\begin{gather}
\underbrace{\forall\textrm{~foldable~} S \in \mathcal{S}^\pqty{k}_\mathrm{cum.}\qc \forall i>0}_{\Downarrow}\nonumber\\
E^{(k+1)}(\Gamma_i,S) \geq E^{(k+1)}(\Gamma_0,S) +\Delta(p_\mathrm{fold},\beta)\,. 
\label{eq:constraints_1}
\end{gather}
Using the notation
\begin{align}
    E^{(k+1)}(\Gamma,S) &=\sum_{nm} \pqty{\sum_{ij}^\prime C_{ij}(\Gamma) \q{i}{n} \q{j}{m}}\, \varepsilon^{(k+1)}_{nm} = \nonumber\\ 
    &= \vb{n}(\Gamma,S) \cdot \boldsymbol{\varepsilon}^{(k+1)},
    \label{eq:constraints_2}
\end{align} 
we compactly rewrite the previous constraints as 
\be \boldsymbol{\varepsilon}^\pqty{k+1}\cdot \vb{x}_i + c_i \geq 0\,, \label{eq:set_constraints}\ee
with suitably chosen arrays $\vb{x}_i$ and constants $c_i$. 
To refine the parameters entering in the designing scoring function so that they minimally violate constraints in Eq.~\eqref{eq:set_constraints}, we resorted to a procedure inspired by Ref.\cite{krauth_mezard_1987}.\\

The refinement procedure starts initializing temporary parameters, $\boldsymbol{\varepsilon}^*$, with $\boldsymbol{\varepsilon}^\pqty{k}$.
At each step, this algorithm uses the temporary parameters to evaluate quantities in Eq.~\eqref{eq:set_constraints}. 
The most violated inequality, weighted by $\eta$, is used to refine the temporary parameters, so that
\be
    \boldsymbol{\varepsilon}^* \rightarrow \boldsymbol{\varepsilon}^* + \eta \, \vb{x}_{i^*},  \qq{with} i^* = \underset{i}{\mathrm{argmin}} \:\boldsymbol{\varepsilon}^* \cdot \vb{x}_i + c_i\,.
\ee

\begin{figure}[H]
    \includegraphics[width=0.9\linewidth]{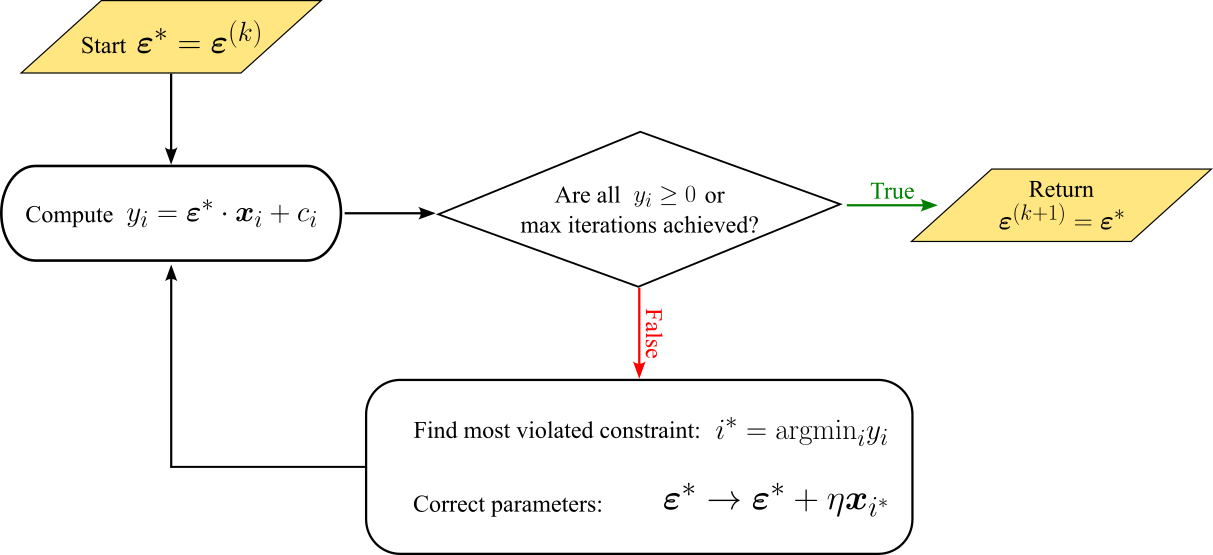}
    \caption{This algorithm takes as inputs parameters entering in the designing function at $k$--th cycle and refines them to minimally violate the set of constraints in Eqs.~\eqref{eq:constraints_1} and~\eqref{eq:constraints_2}. }
    \label{fig:perceptron_algorithm}
\end{figure}

This iterative update proceeds until one of the following conditions occur: (i) the maximum number of iterations is reached or (ii) all constraints are simultaneously satisfied (see Fig.~\ref{fig:perceptron_algorithm}). \\

The number of  constraints that are simultaneously  applied to the parameters of the energy increases at each refinement iteration $k$. Hence, it is convenient to adapt the parameter $\eta$ to decrease at each iteration, leading to finer and finer updates of the energy matrix entries.  In particular, we set $\eta = \frac{\eta_0}{1+3k}$. The numerical values of $\eta_0$ adopted in our simulations are specified in Section S6.

A code implementing this algorithm can be made available upon written request to the authors.

\section*{S6: Numerical details}
\label{sec:numerical_details}

In this section, we provide parameters that are necessary to reproduce the results reported in the main text.

The inverse thermal energy parameter $\beta$ was arbitrarily fixed to $3$ (in appropriate units), the probability threshold defining foldable sequences $p_\mathrm{fold}$ was set to $0.8$, and the ground truth energy maps, for dictionaries $\mathcal{D}_3, \mathcal{D}_4 \mathrm{,\,and\,} \mathcal{D}_5$ are, respectively,
\begin{align*}
	\boldsymbol{\varepsilon}_3 = \mqty(-0.35346 &	0.30399 &	0.42582\\
0.30399 & 	0.17115 &	-0.30167\\
0.42582&	-0.30167 &	0.34102\\) \\\\
\boldsymbol{\varepsilon}_4 = \mqty(0.05375 & 0.21861 & 0.00656 & 0.14191 \\ 0.21861 & 0.43261 & -0.50441 & -0.5146 \\ 0.00656 & -0.50441 & 0.23041 & 0.34485 \\ 0.14191 & -0.5146 & 0.34485 & 0.34976
)\\\\
\boldsymbol{\varepsilon}_5 = \mqty(-0.05777 & 0.26095 & -0.00228 & 0.26162 & 0.0197 \\ 0.26095 & 0.14214 & -0.37257 & 0.13965 & 0.18096 \\ -0.00228 & -0.37257 & 0.04771 & 0.12568 & 0.11891 \\ 0.26162 & 0.13965 & 0.12568 & -0.38521 & 0.02284 \\ 0.0197 & 0.18096 & 0.11891 & 0.02284 & -0.32999)\,.
\end{align*}

Results in Fig.~4a of the main text are derived by designing the target structure in Fig.~2a of the main text employing sequences with compositions $\Bqty{5,5,6},\, \Bqty{5,4,2,5},\,\mathrm{and}\,\Bqty{3,3,2,4,4}$.  
Figure~4b of the main text results when designing targets in Figs.~2a,~2b, and 2c of the main text considering, respectively, sequences with composition $\Bqty{5,5,6},\,\Bqty{7,9,9},\,\mathrm{and}\,\Bqty{12,18,6}$.
The strength of the couplings in the QUBO Hamiltonian defined in Eq.~(8) through~(10) of the main text were set to: $A_1 = 2.1$, $A_2 = 2.1$, and $B = 1$.

The parameter $\eta_0$, introduced in S5, is set to $0.325$ for $\mathcal{D}_3$, to $0.288$ for $\mathcal{D}_4$, and to $0.263$ for $\mathcal{D}_5$. 

\section*{S7: Receiver-Operator Curves}
\label{sec:roc_curves}
In this section, we report the individual receiver-operator curves (ROCs) contributing to the averages shown in Fig.~3 of the main text.
\begin{figure}[H]
    \centering
    \subfloat[]{\includegraphics[width=0.6\linewidth]{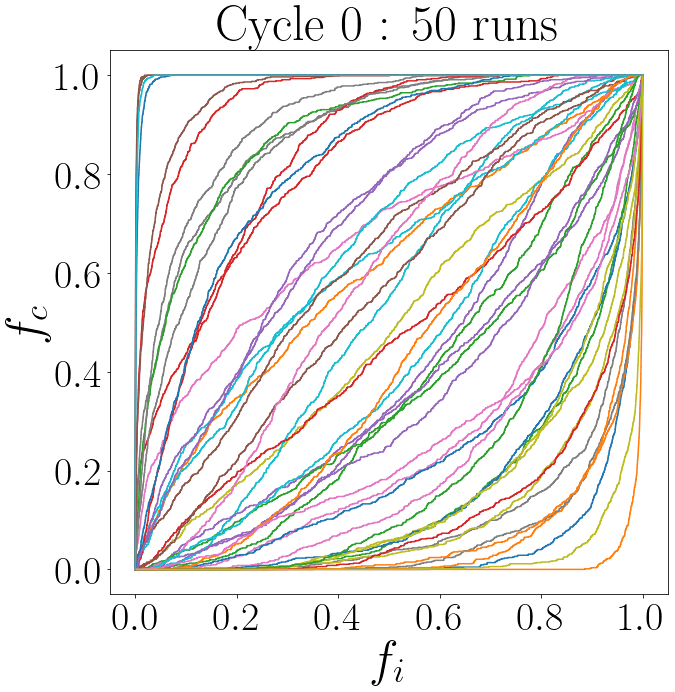}}\\
    \subfloat[]{\includegraphics[width=0.6\linewidth]{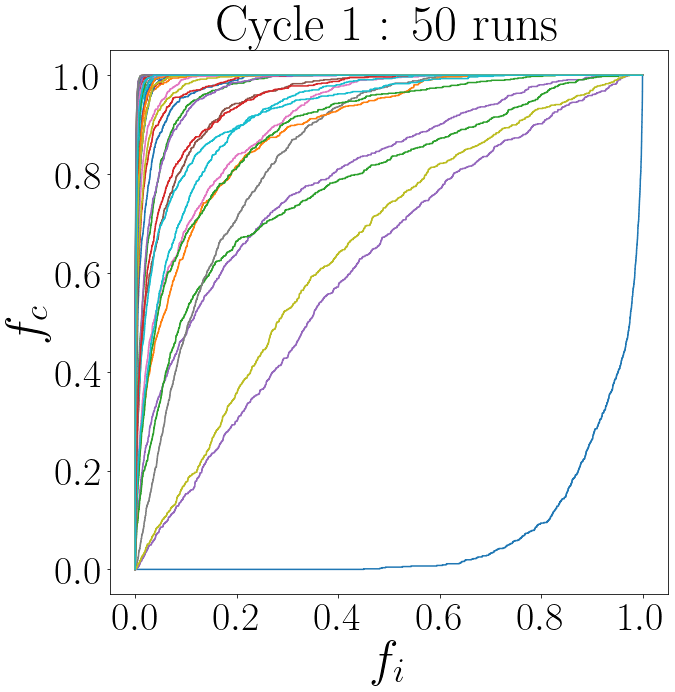}}\\
    \subfloat[]{\includegraphics[width=0.6\linewidth]{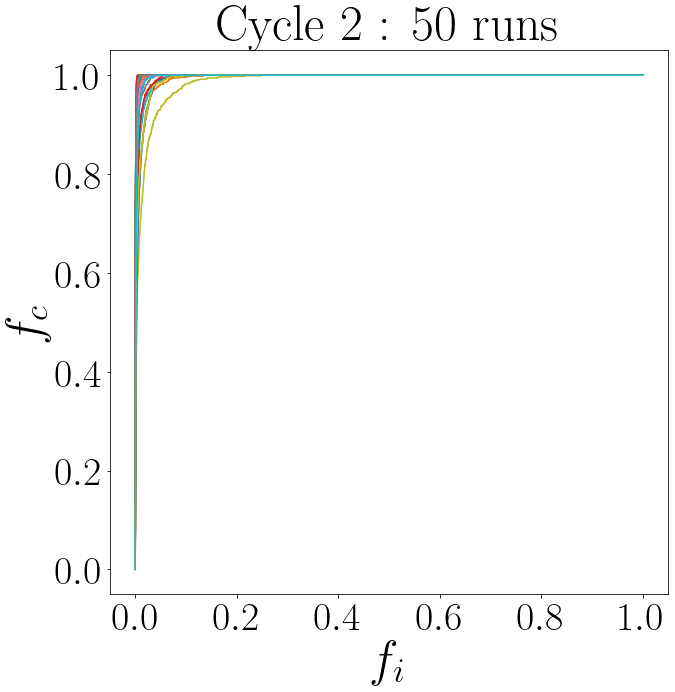}}
    \caption{ROC curves of binary classifiers based on scoring function $G(S)$. }
\end{figure}

\section*{S8: Scaling with respect to problem size}
\begin{figure}[H]
    \centering
    \subfloat[]{\includegraphics[width=0.45\textwidth]{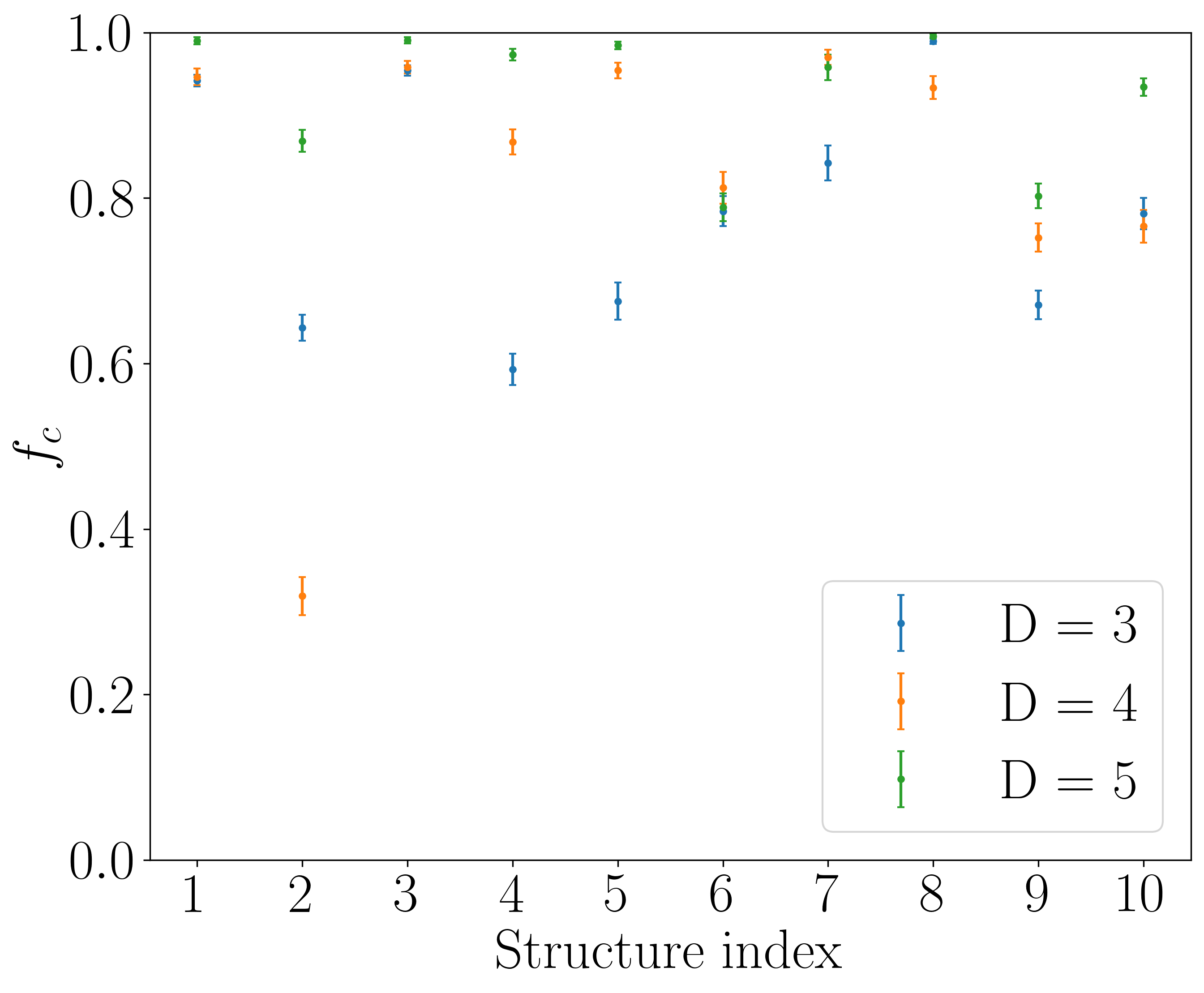}\label{fig:scaling_alhpabet}}\hfill
    \subfloat[]{\includegraphics[width=0.45\textwidth]{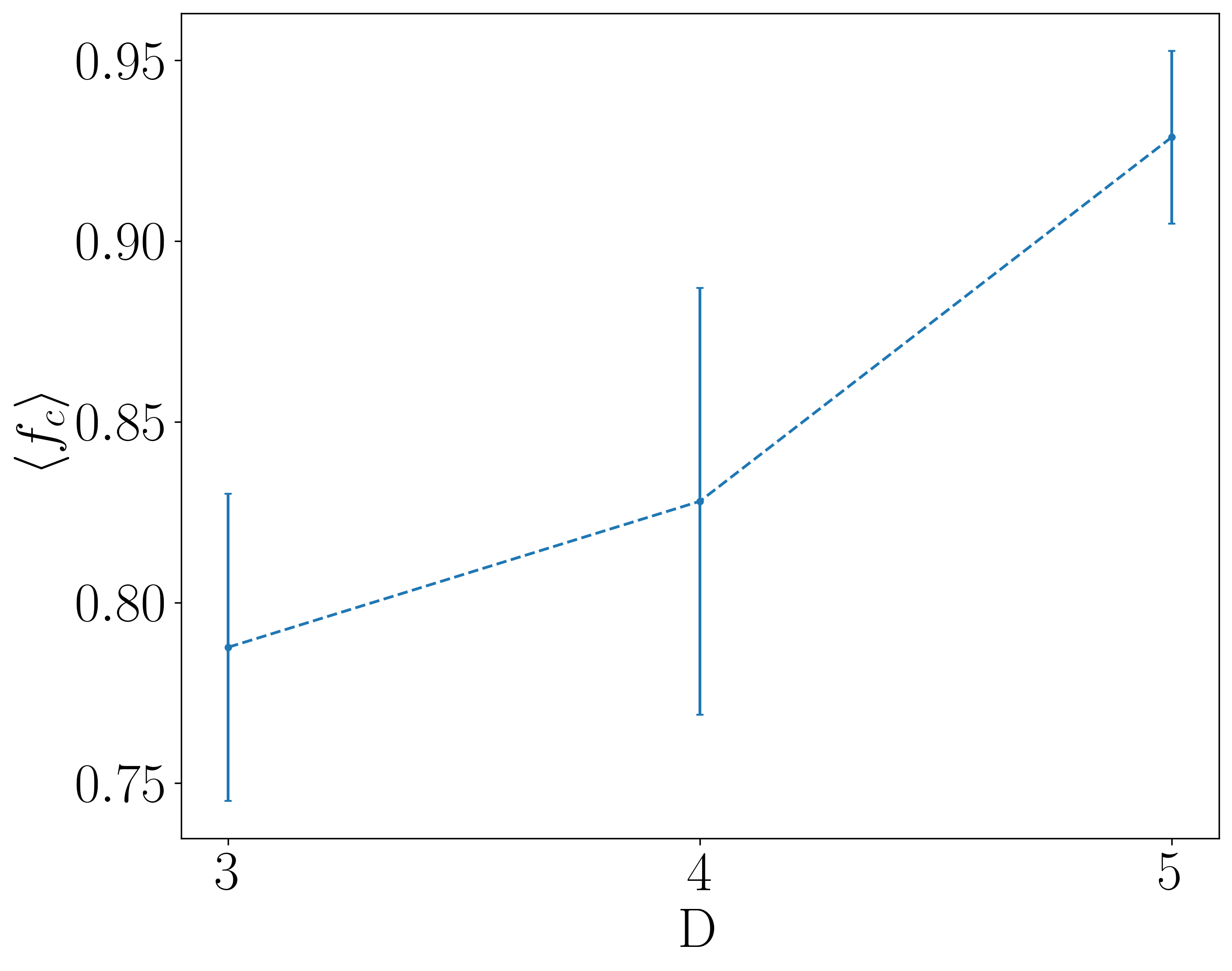}\label{fig:scaling_alphabet:avg}}\\
    \subfloat[]{\includegraphics[width=0.45\textwidth]{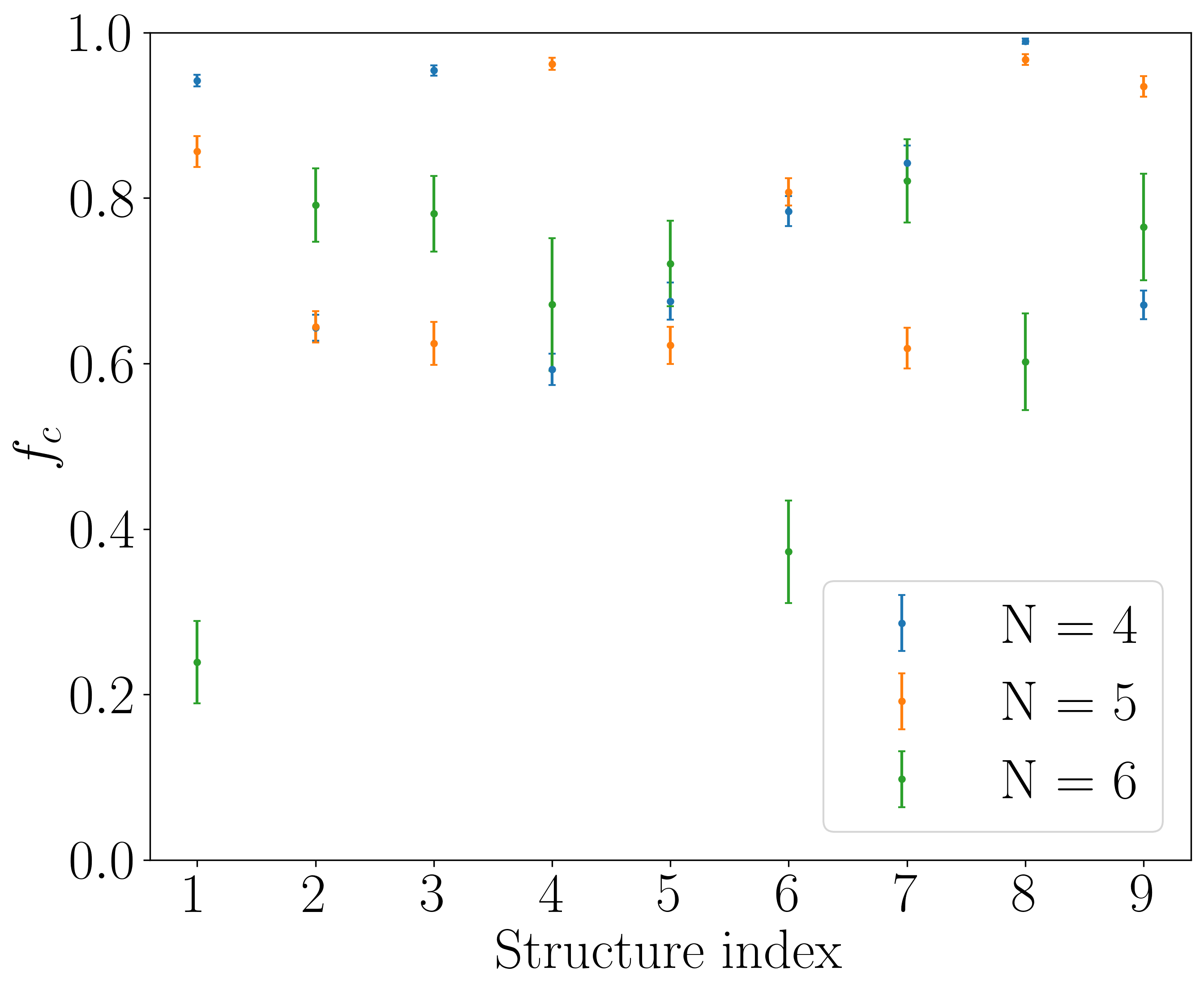}\label{fig:scaling_lattice}}\hfill
    \subfloat[]{\includegraphics[width=0.45\textwidth]{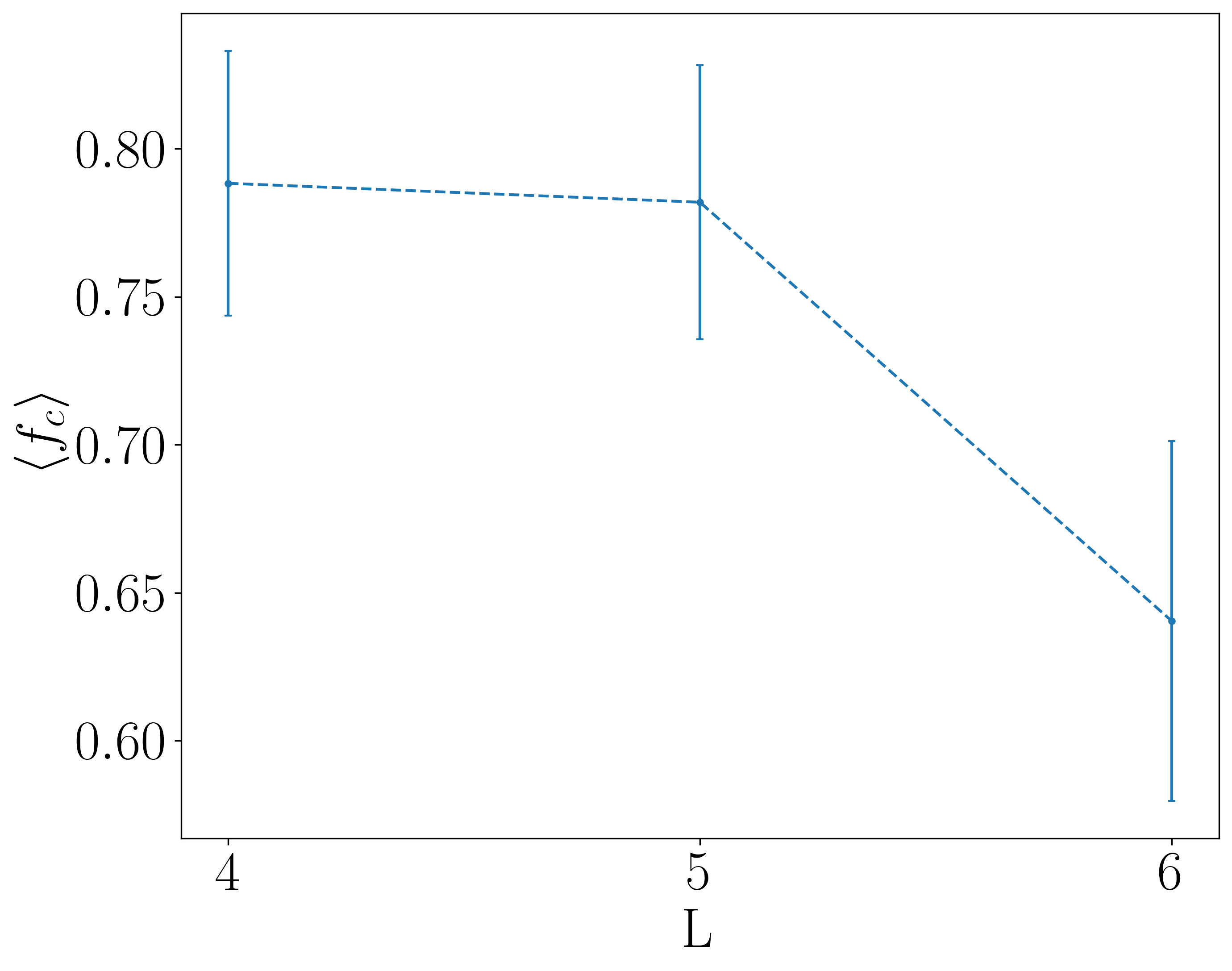}\label{fig:scaling_lattice:avg}}
    \caption{Fraction of sequences correctly folding to the target structure, $f_c$, scaling alphabet size and chain size.  
    \textbf{(a)} We test our algorithm using 10 target structures lying on a $4\times 4$ lattice with a variable alphabet size, $D$. 
    \textbf{(b)} Average values associated with data in (a), indicating that our algorithm improves when considering larger alphabets. 
    \textbf{(c)} Maintaining the alphabet size fixed to 3, we test our algorithm on different lattice sizes, $L\cross L$, where for each lattice format, we select 10 target structures. 
    \textbf{(d)} Average values associated with data in (c), showing that, as the protein size increases, the performance of our algorithm decreases.}
\end{figure}

In this section, we report a study of the success probability of our algorithm as the problem size is increased. 
In Fig.~\ref{fig:scaling_alhpabet} we report the success rates associated with the design of 10 target structures, embedded on a $4\times 4$ lattice, using alphabets composed of 3, 4, and 5 letters. Average values, presented in Fig.~\ref{fig:scaling_alphabet:avg}, indicate that larger alphabets improve the overall success rate of our algorithm.
In Fig.~\ref{fig:scaling_lattice}, we report the success rate of our algorithm in designing a set of 30 structures, evenly distributed across $4\times 4$, $5\times 5$, and $6\times 6$ lattices, using an alphabet of 3 letters. Averaging these results over the structures' ensemble, see Fig.~\ref{fig:scaling_lattice:avg}, shows that performances are negatively affected when the size of the protein increases. \ 

\onecolumngrid\
\twocolumngrid\
\end{document}